\documentclass[a4paper,fleqn]{cas-dc}

\usepackage[numbers]{natbib}
\usepackage{xcolor}
\usepackage{tikz}
\usepackage{subcaption}
\usetikzlibrary{positioning,decorations.pathreplacing,arrows.meta,calc}
\usepackage{listings}
\lstdefinestyle{hmcconfig}{
    basicstyle=\ttfamily\footnotesize,
    columns=fullflexible,
    breaklines=true,
    showstringspaces=false,
    commentstyle=\itshape\color{gray},
    keywordstyle=\bfseries,
    morecomment=[l]{\#},
    frame=single,
    framesep=4pt,
    xleftmargin=4pt,
    xrightmargin=4pt,
    captionpos=b
}

\begin{document}
\let\WriteBookmarks\relax

\shorttitle{HybridMC: fast simulation for LHCb ECAL Upgrade~II}

\shortauthors{M.~Pizzichemi \textit{et al.}}

\title [mode = title]{HybridMC: a fast detailed simulation framework for optical calorimeter technologies at the LHCb ECAL Upgrade~II}

\author[1,2]{Marco Pizzichemi}[orcid=0000-0001-5189-230X]%
\author[3]{Liupan An}%
\author[4]{Federico Betti}%
\author[5,2]{Aleksandar Bordelius}%
\author[2]{Loris Martinazzoli}%
\author[2]{Philipp Roloff}%
\author[1,2]{Matteo Salomoni}%
\author[3]{Yanxi Zhang}%

\affiliation[1]{organization={Universit\`a degli Studi di Milano-Bicocca},
            city={Milano},
            country={Italy}}
\affiliation[2]{organization={CERN},
              city={Geneva},
              country={Switzerland}}
\affiliation[3]{organization={Peking University},
              city={Beijing},
              country={China}}
\affiliation[4]{organization={INFN Sezione di Bologna},
              city={Bologna},
              country={Italy}}
\affiliation[5]{organization={Karlsruhe Institute of Technology},
              city={Karlsruhe},
              country={Germany}}

\cortext[1]{Corresponding author}

\begin{abstract}
The detailed Monte Carlo simulation of complex optical calorimeters is dominated by the generation and tracking of scintillation and Cherenkov photons, which in standard Geant4 ray-tracing inflates the CPU time per electromagnetic shower by up to three orders of magnitude compared to simulating only the energy deposition. At the luminosity and detector granularity foreseen for the LHCb ECAL Upgrade~II, this cost becomes prohibitive for both detector R\&D and large-scale production of simulated samples. The optical photons cannot simply be discarded, because the time resolution of a few tens of picoseconds targeted by the upgrade, essential to mitigate the pile-up, is driven by their arrival-time distribution at the photo-detectors. This paper presents the HybridMC framework, a fast and detailed simulation toolkit dedicated to optical calorimeters that decouples shower development from optical transport. Energy deposition is handled by a standard Geant4 simulation in which the optical photons are not propagated, while the light transport is reproduced by a parametrised \emph{summary Monte Carlo} step based on probability density functions extracted from a one-time full ray-tracing optical calibration. The framework has been validated through a Monte Carlo-to-Monte Carlo procedure, showing good statistical agreement with full ray-tracing for all module types of the upgraded LHCb ECAL, from the arrival times of the individual photons up to the reconstructed energy and time resolutions. The corresponding gain in CPU time amounts to more than two orders of magnitude, with module-dependent acceleration factors ranging from 150 to 480. A comparison to test-beam data collected at DESY~II and at the CERN SPS shows that the description of a real prototype implemented in the framework reproduces the momentum dependence of the measured energy and time resolutions, up to a small residual constant term. HybridMC retains the microscopic optical detail that is not accessible to ultra-fast flash simulators based on generative models, at a CPU cost compatible with large-scale production on standard computing infrastructure.
\end{abstract}

\begin{highlights}
\item Fast Monte Carlo for optical calorimeters: Geant4 shower + parametrised optical transport.
\item Speed-up of $150\times$ to $480\times$ over full Geant4 ray-tracing of optical photons.
\item MC-to-MC validation: Kolmogorov-Smirnov $p>0.05$ for more than $93\%$ of simulated pulses.
\item Reproduces the test-beam energy and time resolution of a SpaCal prototype module.
\end{highlights}


\maketitle

\section{Introduction}
\label{introduction}

The upcoming High-Luminosity Large Hadron Collider (HL-LHC) era presents high-energy physics (HEP) experiments with computational and experimental challenges of unprecedented magnitude. Among the experiments installed on the CERN LHC accelerator, the LHCb detector is a single-arm forward spectrometer, covering a pseudorapidity range of $2<\eta<5$ and specifically designed for the precise study of particles containing beauty (b) and charm (c) quarks \cite{The_LHCb_Collaboration_2008}. The current LHCb Electromagnetic Calorimeter (ECAL) is a large array of Shashlik sampling modules covering approximately 50 square metres, located downstream of the experiment's dipole magnet. It is primarily optimised for the identification and transverse energy measurement of electrons, photons, and neutral pions ($\pi^0$) in the energy range of a few GeV to 100 GeV \cite{calo1_2000,calo2_2020}.

\begin{table*}[pos=tbp]
\caption{Main parameters of the module configurations used as case studies in this paper. The first three are SpaCal modules and the last is a Shashlik module. For the SpaCal modules the fibre column gives the cross-section and the transverse pitch of the scintillating fibres. For the Shashlik module it gives the diameter and pitch of the wavelength-shifting (WLS) readout fibres, the active medium being $4$~mm polystyrene tiles interleaved with $2$~mm lead plates. ``Fibres per cell'' is the number of scintillating (SpaCal) or WLS (Shashlik) fibres read out by a single cell. ``Cells per module'' is the transverse readout granularity and ``cell size'' the transverse extent of a cell. The Shashlik module is read out as a single cell per section, collecting all of its WLS fibres. ``Sections'' is the number of independently read longitudinal segments, with their depths in parentheses. All modules have a transverse size of about $120\times120$~mm$^2$.}\label{table:configurations}
\begin{tabular*}{\tblwidth}{@{}LCCCCCCC@{}}
\toprule
\textbf{Configuration} & \textbf{Absorber} & \textbf{Scintillator} & \textbf{Fibre (size, pitch)} & \textbf{Fibres/cell} & \textbf{Cells/module} & \textbf{Cell size} & \textbf{Sections} \\
\midrule
W-GAGG        & Tungsten & GAGG        & $1\times1$~mm$^2$, $1.67$~mm      & $81$  & $64$ ($8\times8$)  & $15.03$~mm & $2$ ($50+100$~mm)    \\
Pb-Polystyrene & Lead    & Polystyrene & $\varnothing1.5$~mm, $2.16$~mm    & $196$ & $16$ ($4\times4$)  & $30.24$~mm & $2$ ($100+200$~mm)  \\
W-Polystyrene & Tungsten & Polystyrene & $1\times1$~mm$^2$, $1.67$~mm      & $144$ & $36$ ($6\times6$)  & $20.04$~mm & $1$ ($200$~mm)      \\
Shashlik      & Lead     & Polystyrene & $\varnothing1$~mm WLS, $15.25$~mm & $64$ & $1$ & $120$~mm & $2$ ($19+48$ layers) \\
\bottomrule
\end{tabular*}
\end{table*}

For the LHCb experiment, the planned Upgrade II will operate at an instantaneous luminosity of 1.0 $\times$ 10$^{34}$ cm$^{-2}$s$^{-1}$, resulting in a dense environment with an average of about 30 proton-proton interactions per bunch crossing. To mitigate this massive pile-up and preserve physics performance, the upgraded ECAL must feature increased granularity and exceptional time resolution on the order of tens of picoseconds. Proposed solutions include highly segmented technologies such as Spaghetti Calorimeter (SpaCal)~\cite{SpaCal_proposal_1987} modules equipped with bright scintillating crystal fibres or polystyrene fibres \cite{collaboration2021framework}. While these advanced detector designs are crucial for future physics discoveries, they severely exacerbate an already critical computing challenge. Detailed Monte Carlo (MC) simulations, primarily driven by the Geant4 toolkit~\cite{agostinelli2003geant4}, currently consume approximately 90\% of the LHCb experiment's total CPU budget \cite{LHCb_Lamarr_2024}. The ECAL alone accounts for up to 60\% of this simulation time in Run~3 beam conditions and geometry, due to the dense proliferation of secondary particles within electromagnetic showers \cite{cds_fast_simulation_2021,cds_michal_2021}.

The primary bottleneck in simulating highly granular optical calorimeters is the generation and propagation of optical photons (scintillation and Cherenkov radiation). Bright scintillators generate tens of thousands of photons per MeV of deposited energy. In standard Geant4 full ray-tracing simulations, each of these optical photons must be individually tracked through complex boundaries, requiring the evaluation of Fresnel refractions, total internal reflection, absorption, and surface roughness via formalisms like the UNIFIED model~\cite{Levin1996Unified}.

Introducing optical photon tracking can dramatically inflate the simulation time of a single electromagnetic shower, rendering the process computationally prohibitive. To quantify the computational burden imposed by light transport, we report here the results of a simple benchmark test conducted by repeatedly simulating a single 1 GeV/c electron incident on the centre of a SpaCal module, in which a tungsten absorber (120$\times$120$\times$200 mm$^3$) is embedded with 5184 scintillating fibres, each 1.0$\times$1.0$\times$200 mm$^3$.
Total CPU times were recorded under two distinct configurations: one restricting the simulation to purely electromagnetic shower development by disabling optical photons, and another enabling full optical photon production and ray-tracing. These tests were executed using version 11 of the Geant4 toolkit on a single core of an Intel Core i7-10750H processor. When simulating only the electromagnetic shower, the average computation time per 1 GeV/c incident electron was measured at approximately 0.35 s. However, incorporating the generation and step-by-step tracking of optical photons drastically inflated the processing time to an average of 880 s. Consequently, the generation and tracking of optical photons introduce a computational penalty of a factor of 2500. This result does not represent a precise quantification of the expected CPU time increase for all simulations that include optical photons, as the actual factor depends on the specific module characteristics (technology, geometry, materials, surface states, etc.), as well as on the physics-list used, but it illustrates the scale of the problem.

The electromagnetic particles measured by the LHCb calorimeter typically carry energies from $1$ to $100$~GeV, reaching a few hundred GeV only in the central region closest to the beam. Since the CPU cost scales linearly with the deposited energy, simulating this range with full optical ray-tracing becomes prohibitive, particularly during the R\&D phase, where numerous detector configurations must be iteratively simulated and evaluated. On the other hand, bypassing optical photons entirely is not a viable option during the detector R\&D and optimisation phases, as the detailed modelling of light transport is essential, among other things, for predicting time resolution, evaluating the impact of scintillator surface states (polishing/depolishing), and assessing performance degradation due to radiation damage.

Current strategies to accelerate calorimeter simulations generally fall into two broad categories: ultra-fast deep generative models and hardware-accelerated ray-tracing. Machine-learning-based flash simulations using Generative Adversarial Networks (GANs) and Variational Auto-Encoders (VAEs), such as Lamarr and the CaloML frameworks, bypass Geant4 entirely to directly parametrise the high-level energy deposition of showers \cite{cds_michal_2025,chekalina2019generative}. While these achieve speedups of up to five orders of magnitude, they model macroscopic detector responses and are less suited for hardware design iterations where internal optical properties are modified. Alternatively, hardware-accelerated approaches offload optical photon propagation to Graphics Processing Units (GPUs) using tools like Opticks \cite{Opticks} (which translates Geant4 geometries to NVIDIA OptiX) or the Mitsuba3 rendering engine \cite{davis2023mitsuba3}. While these solutions provide impressive speed-ups (e.g. up to $200\times$--$1500\times$), they rely on specific heterogeneous computing architectures and present complex integration challenges within existing, heavily CPU-based distributed computing grids.

To bridge the gap between microscopic optical precision and computational feasibility, this paper presents the HybridMC framework. HybridMC is a fast, detailed simulation toolkit dedicated to calorimeters that decouples shower development from optical light transport. It operates by using Geant4 exclusively for particle interactions and energy deposition, while completely forbidding the costly Geant4-based propagation of optical photons. Instead, the optical transport is reproduced using a parametrised \emph{summary Monte Carlo} approach based on pre-computed lookup tables of probability density functions. By encoding the complex physics of extraction efficiency and time distributions from a one-time optical calibration, the HybridMC framework retains the necessary physical accuracy regarding surface states and geometry while accelerating the simulation by factors ranging from 150 to 480 compared to full ray-tracing.

The remainder of this paper is organised as follows. Section~\ref{sec:framework} describes the HybridMC architecture and its constituent stages, from the one-time optical calibration to the event-by-event simulation pipeline. Section~\ref{sec:validation} presents the Monte Carlo-to-Monte Carlo validation against full Geant4 ray-tracing, carried out on one representative module per technology and spanning from the arrival times of the individual photons up to the reconstructed energy and time resolutions. Section~\ref{sec:gain} reports the CPU-time gain achieved on the module technologies relevant to the LHCb ECAL Upgrade~II. Section~\ref{sec:comparison} complements the validation with a comparison to test-beam data collected on a SpaCal prototype, demonstrating that the modelling of a real module within the framework is adequate. Section~\ref{sec:discussion} discusses the validity domain of the parametrisation and positions HybridMC within the landscape of accelerated calorimeter simulations, and Section~\ref{sec:conclusions} draws the conclusions.

Throughout this paper, the framework is exercised on a set of module configurations representative of the technologies under consideration for the LHCb ECAL Upgrade~II, whose main geometrical and material parameters are summarised in Table~\ref{table:configurations}. Three are SpaCal modules, differing in the absorber (tungsten or lead) and in the scintillating fibre (GAGG crystal or polystyrene), and one is a 1-cell Shashlik module. The W-GAGG SpaCal and the Shashlik modules carry the full Monte Carlo-to-Monte Carlo validation of Section~\ref{sec:validation}, while all four enter the CPU-time study of Section~\ref{sec:gain}.

\section{The HybridMC framework}\label{sec:framework}

HybridMC has been developed under four design requirements that follow from the use cases described in Section~\ref{introduction}:
\begin{itemize}
\item \emph{Optical fidelity:} the simulation must reproduce the physics of optical photon transport, i.e. surface roughness, total internal reflection, reflector materials, optical coupling. This is essential for R\&D iterations on detector technologies and is discarded by ultra-fast flash simulators.
\item \emph{CPU cost compatible with large-scale studies:} the framework must run on the existing CPU-based distributed computing grid of the LHC experiments, without relying on heterogeneous hardware accelerators.
\item \emph{Modular and reconfigurable:} all module-specific parameters, i.e. geometry, materials, surface states, readout layout, must be controlled through plain-text configuration files, so that new prototype variants can be studied without recompiling the framework. Since the executable does not need to be adapted to a specific module type, this also makes it possible to simulate different module types at the same time, within the same run.
\item \emph{Native multi-technology support:} both SpaCal and Shashlik modules and arbitrary mixtures thereof must be handled at the full-ECAL scale, following the heterogeneous design of the LHCb ECAL Upgrade~II.
\end{itemize}

\begin{figure*}[pos=tbp]
\centering
\begin{tikzpicture}[
    font=\small,
    >=Latex,
    box/.style={
        rectangle, rounded corners=2pt,
        draw, line width=0.6pt,
        minimum height=14mm, minimum width=34mm,
        align=center, inner sep=3pt
    },
    prep/.style={box, fill=blue!10,   draw=blue!60!black},
    geant/.style={box,fill=orange!18, draw=orange!70!black},
    hmcS/.style={box, fill=green!18,  draw=green!55!black},
    digi/.style={box, fill=red!12,    draw=red!60!black},
    flow/.style={->, thick, shorten >=1pt, shorten <=1pt},
    tier/.style={font=\scriptsize\itshape, gray!70!black}
]

\node[prep] (cal) at (0,0) {Optical Calibration\\[1pt]\scriptsize\textit{Geant4 + HybridMC}};

\node[font=\scriptsize, anchor=east, align=right, blue!45!black] at ([xshift=-6mm]cal.west)
   {summarises optical\\transport};
\node[font=\scriptsize, anchor=west, align=left,  blue!45!black] at ([xshift=6mm]cal.east)
   {\textbf{only once}\\per module type};

\node[geant, below=22mm of cal, xshift=-42mm] (edep) {Energy deposition\\[1pt]\scriptsize\textit{Geant4}};
\node[hmcS,  below=22mm of cal]               (xpt)  {Parametrised\\optical transport\\[1pt]\scriptsize\textit{HybridMC}};
\node[digi,  below=22mm of cal, xshift=42mm]  (dig)  {Digitisation and\\signal formation\\[1pt]\scriptsize\textit{HybridMC}};

\draw[flow] (cal.south) -- node[yshift=2mm,font=\scriptsize, right=1pt]{\textsc{PDF}s} (xpt.north);
\draw[flow] (edep.east) -- (xpt.west);
\draw[flow] (xpt.east)  -- (dig.west);

\draw[flow] ([xshift=-12mm]edep.west) -- node[above,font=\scriptsize]{primary} node[below,font=\scriptsize]{event} (edep.west);
\draw[flow] (dig.east) -- node[xshift=1mm,yshift=2mm,font=\scriptsize]{observables} ++(12mm,0);

\node[tier, rotate=90, anchor=south] at ([xshift=-72mm]cal.west) {Preparatory};
\node[tier, rotate=90, anchor=south] at ([xshift=-30mm]edep.west) {MC production};

\draw[dash pattern=on 2pt off 2pt, gray!50]
   ([xshift=-78mm,yshift=-11mm]cal.south west) -- ([xshift=12mm,yshift=+11mm]dig.north east);

\end{tikzpicture}
\caption{Schematic representation of the HybridMC framework. The pipeline is composed of a one-time preparatory Optical Calibration stage followed by an event-by-event chain of an Energy Deposition step, a Parametrised Optical Transport step, and a Digitisation and Signal Formation step. The Geant4-based stages handle the particle shower and the calibration ray-tracing, while the parametrised stages, implemented as standalone C++ programs, reproduce the optical photon transport from the pre-computed probability density functions (\textsc{pdf}s) produced once per module type by the calibration stage.}\label{fig:hmc_structure}
\end{figure*}
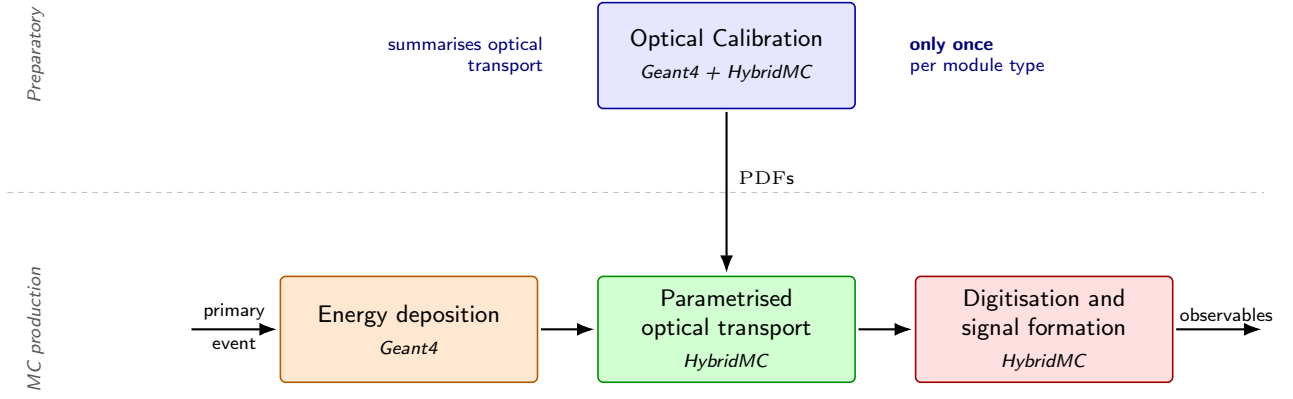
The architecture of the framework (see Figure~\ref{fig:hmc_structure}) is structured as a modular pipeline split into consecutive steps, comprising a one-time preparatory phase followed by a multi-stage event-by-event simulation. The preparatory phase consists of the \emph{optical calibration}, a specialised full ray-tracing simulation performed only once per module type, which characterises the light transport properties of the specific calorimeter geometry and encodes them into lookup tables of probability density functions (PDFs). The event-by-event simulation begins with the \emph{energy deposition} step, a standard Geant4-based simulation that handles the full particle interaction and shower development to produce a 3D energy deposition map. During this step, the Geant4 generation and propagation of optical photons are deactivated to save CPU time. The core acceleration phase, named \emph{parametrised optical transport}, follows: a custom C++ algorithm iterates over the Geant4 energy deposition map and uses the pre-computed calibration PDFs to parametrically determine the detection probability and arrival time of the optical photons. The final \emph{digitisation and signal formation} step bridges the gap between simulated photons and detector electronics by converting the detected photons into realistic, digitised electronic waveforms, accounting for quantum efficiency and transit time spread, from which the reconstructed energy and time are then extracted.

A more in-depth description of the methodologies, internal algorithms, and implementations of each of the constituent steps reported above is provided in the following sections. Since the first two steps are both performed using the same Monte Carlo simulation, based on the Geant4 simulation toolkit, this simulation will first be described.

\subsection{Structure of the Monte Carlo simulation}\label{sec:mcsim}

\begin{figure}[pos=tbp]
\centering
\includegraphics[width=0.7\linewidth]{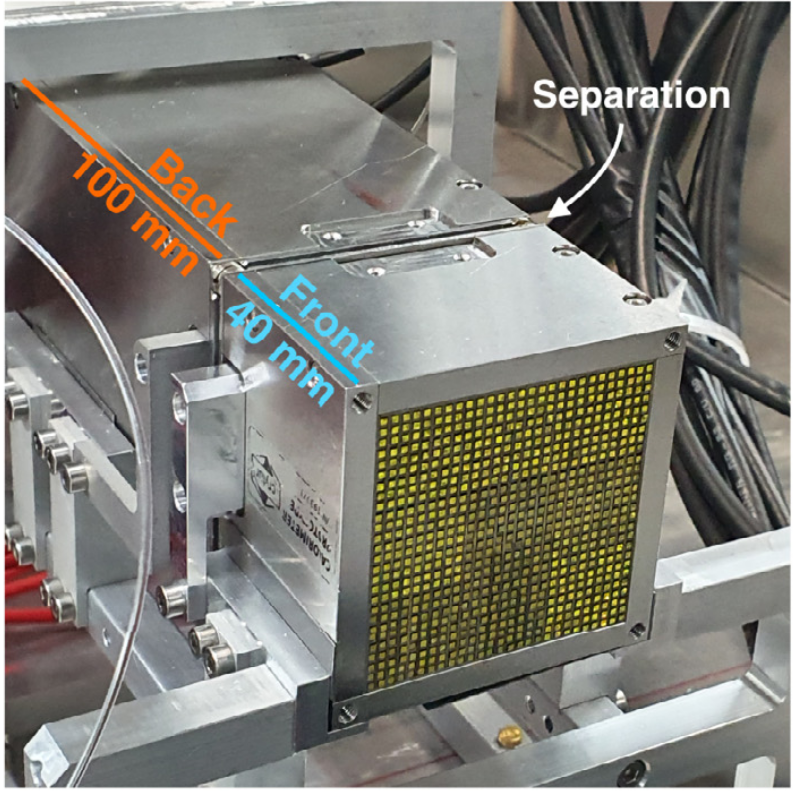}
\caption{Photograph of a SpaCal prototype module used during the LHCb ECAL Upgrade~II R\&D campaign. Reproduced from \cite{An_2022_TB}.}
\label{fig:photo_spacal}
\end{figure}

\begin{figure}[pos=tbp]
\centering
\includegraphics[width=0.7\linewidth]{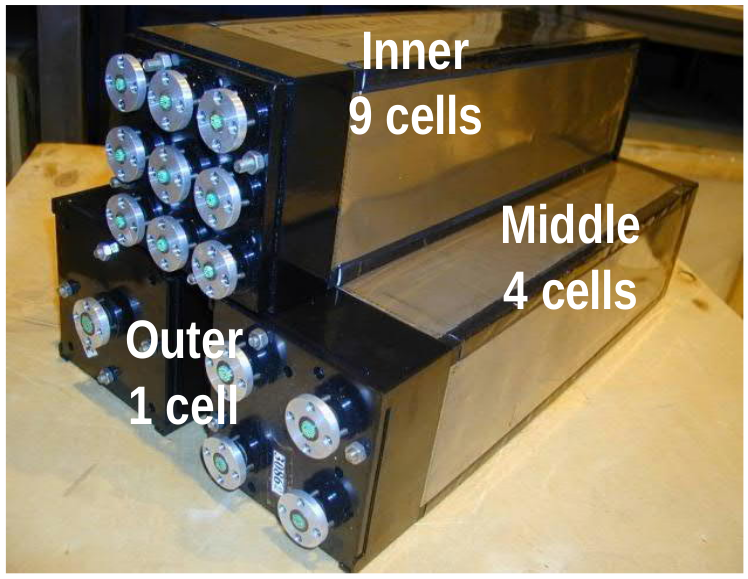}
\caption{Photograph of the three types of Shashlik modules used in the current LHCb ECAL: 1-cell, 4-cell, and 9-cell configurations.}
\label{fig:photo_shashlik}
\end{figure}

Given its application to the LHCb ECAL Upgrade II, the HybridMC must be able to simulate calorimeter modules based on both SpaCal and Shashlik technologies, shown in Figures~\ref{fig:photo_spacal} and~\ref{fig:photo_shashlik}. The geometrical modelling of both technologies is implemented in the framework, giving the user the possibility to decide the features of the modules via a simple configuration file. %

The framework leaves the user free to select the Geant4 physics list that best suits the application. In the LHCb ECAL Upgrade~II studies reported here, the \texttt{FTFP\_BERT} hadronic physics list with the Electromagnetic Option Opt0 has been adopted. Range production cuts for electrons, positrons, photons, and hadrons are set globally through the \texttt{defaultCut} configuration key. Throughout this work the cut is fixed at $0.025$~mm to preserve the maximum spatial precision in the development of the electromagnetic shower and in the energy deposition pattern inside the scintillating elements.

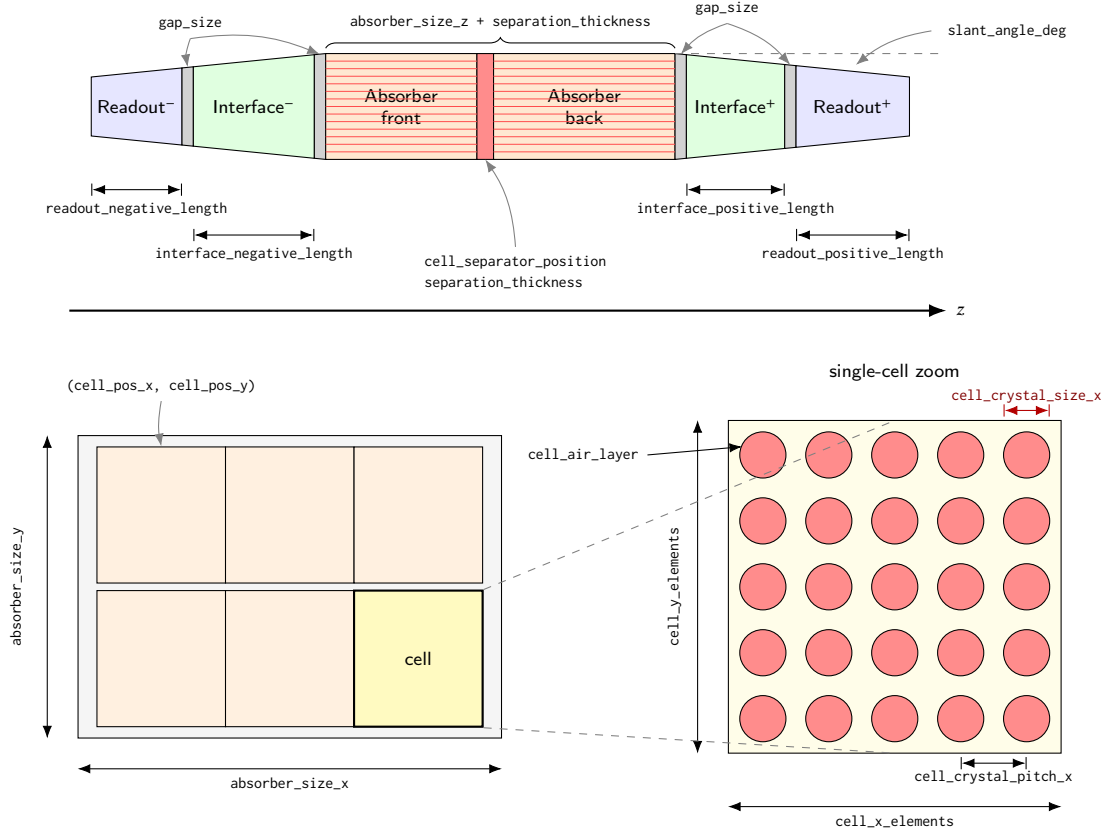
\begin{figure*}[pos=tbp]
\centering
\begin{tikzpicture}[
    font=\scriptsize,
    >=Latex,
    x=1mm, y=1mm,
    vol/.style={draw},
    ann/.style={font=\scriptsize\ttfamily}
]

\draw[vol,fill=blue!10] (-54.1,-3.90) -- (-42.1,-5.10) -- (-42.1,5.10) -- (-54.1,3.90) -- cycle;
\node at (-48.1,0) {Readout$^-$};
\draw[vol,fill=gray!35] (-42.1,-5.10) -- (-40.6,-5.25) -- (-40.6,5.25) -- (-42.1,5.10) -- cycle;
\draw[vol,fill=green!12] (-40.6,-5.25) -- (-24.6,-6.85) -- (-24.6,6.85) -- (-40.6,5.25) -- cycle;
\node at (-32.6,0) {Interface$^-$};
\draw[vol,fill=gray!35] (-24.6,-6.85) -- (-23.1,-7) -- (-23.1,7) -- (-24.6,6.85) -- cycle;
\draw[vol,fill=orange!18] (-23.1,-7) rectangle (-3.1,7);
\node[align=center] at (-13.1,0) {Absorber\\front};
\draw[vol,fill=red!45] (-3.1,-7) rectangle (-0.9,7);
\draw[vol,fill=orange!18] (-0.9,-7) rectangle (23.1,7);
\node[align=center] at (11.1,0) {Absorber\\back};
\draw[vol,fill=gray!35] (23.1,-7) -- (24.6,-6.85) -- (24.6,6.85) -- (23.1,7) -- cycle;
\draw[vol,fill=green!12] (24.6,-6.85) -- (37.6,-5.55) -- (37.6,5.55) -- (24.6,6.85) -- cycle;
\node at (31.1,0) {Interface$^+$};
\draw[vol,fill=gray!35] (37.6,-5.55) -- (39.1,-5.40) -- (39.1,5.40) -- (37.6,5.55) -- cycle;
\draw[vol,fill=blue!10] (39.1,-5.40) -- (54.1,-3.90) -- (54.1,3.90) -- (39.1,5.40) -- cycle;
\node at (46.6,0) {Readout$^+$};

\foreach \yy in {-6,-5,...,6}{
  \draw[red!70,line width=0.3pt] (-23.1,\yy) -- (-3.1,\yy);
  \draw[red!70,line width=0.3pt] (-0.9,\yy)  -- (23.1,\yy);
}

\draw[dashed,gray] (23.1,7) -- (58,7);
\draw[->,thin,gray] (57,10) to[bend right=15] (47,5.5);
\node[ann,right] at (58,10) {slant\_angle\_deg};

\draw[->,thick] (-57,-27) -- (59,-27) node[right]{$z$};

\draw[|<->|,thin] (-54.1,-11) -- (-42.1,-11);
\node[ann,below] at (-48.1,-11.5) {readout\_negative\_length};
\draw[|<->|,thin] (24.6,-11) -- (37.6,-11);
\node[ann,below] at (31.1,-11.5) {interface\_positive\_length};

\draw[|<->|,thin] (-40.6,-17) -- (-24.6,-17);
\node[ann,below] at (-32.6,-17.5) {interface\_negative\_length};
\draw[|<->|,thin] (39.1,-17) -- (54.1,-17);
\node[ann,below] at (46.6,-17.5) {readout\_positive\_length};

\draw[decorate,decoration={brace,amplitude=4pt,raise=2pt}]
    (-23.1,7) -- (23.1,7)
    node[midway,above=6pt,ann]{absorber\_size\_z + separation\_thickness};

\node[ann,align=left] (slab) at (2,-22) {cell\_separator\_position\\separation\_thickness};
\draw[->,thin,gray] (slab.north) to[bend right=15] (-2,-7);

\node[ann,align=left] (glabl) at (-41,11) {gap\_size};
\draw[->,thin,gray] (glabl.south) to[bend right=15] (-41.35,5.2);
\draw[->,thin,gray] (glabl.south) to[bend left=15] (-23.85,6.9);

\node[ann,align=left] (glabr) at (30,13) {gap\_size};
\draw[->,thin,gray] (glabr.south) to[bend left=15] (38.35,5.5);
\draw[->,thin,gray] (glabr.south) to[bend right=15] (23.85,6.9);

\end{tikzpicture}

\vspace{4mm}

\begin{tikzpicture}[font=\scriptsize,>=Latex,x=1mm,y=1mm]
\node[draw,fill=gray!8,minimum width=56mm,minimum height=40mm] (abox) at (0,0) {};
\foreach \i in {0,1,2}{
  \foreach \j in {0,1}{
    \node[draw,fill=orange!12,minimum width=17mm,minimum height=18mm]
      (c-\i-\j) at ({-17+\i*17}, {-9.5+\j*19}) {};
  }
}
\node[draw,thick,fill=yellow!30,minimum width=17mm,minimum height=18mm]
  (hcell) at (17,-9.5) {};
\node[font=\scriptsize] at (17,-9.5) {cell};

\draw[<->,thin] ($(abox.south west)+(0,-4mm)$) -- ($(abox.south east)+(0,-4mm)$);
\node[font=\scriptsize\ttfamily,below=4mm of abox]{absorber\_size\_x};
\draw[<->,thin] ($(abox.south west)+(-4mm,0)$) -- ($(abox.north west)+(-4mm,0)$);
\node[font=\scriptsize\ttfamily,rotate=90,anchor=south]
    at ($(abox.west)+(-6mm,0)$){absorber\_size\_y};

\node[font=\scriptsize\ttfamily,above=6mm of c-0-1] (cpos) {(cell\_pos\_x, cell\_pos\_y)};
\draw[->,thin,gray] (cpos.south) to[bend right=10] (c-0-1.north);

\begin{scope}[xshift=80mm,x=10.9mm,y=10.9mm]
  \node[draw,fill=yellow!10,minimum width=44mm,minimum height=44mm] (zoom) at (0,0) {};
  \foreach \i in {0,...,4}{
    \foreach \j in {0,...,4}{
      \draw[fill=red!45,draw=black,thin] ({-1.6+\i*0.8}, {-1.6+\j*0.8}) circle (0.28);
    }
  }
  \node[font=\scriptsize,above=4mm of zoom]{single-cell zoom};

  \draw[|<->|,thin] (0.8,-2.14) -- (1.6,-2.14);
  \node[font=\scriptsize\ttfamily,below] at (1.2,-2.13) {cell\_crystal\_pitch\_x};

  \draw[|<->|,red!70!black,line width=0.4pt] (1.32,2.14) -- (1.88,2.14);
  \node[font=\scriptsize\ttfamily,above,red!50!black] at (1.6,2.13) {cell\_crystal\_size\_x};

  \draw[->,thin] (-3,1.6) -- (-1.85,1.7);
  \node[font=\scriptsize\ttfamily,left,align=right] at (-3,1.6) {cell\_air\_layer};

  \draw[<->,thin] ($(zoom.south west)+(0,-7mm)$) -- ($(zoom.south east)+(0,-7mm)$);
  \node[font=\scriptsize\ttfamily,below=7mm of zoom]{cell\_x\_elements};
  \draw[<->,thin] ($(zoom.south west)+(-4mm,0)$) -- ($(zoom.north west)+(-4mm,0)$);
  \node[font=\scriptsize\ttfamily,rotate=90,anchor=south]
      at ($(zoom.west)+(-5mm,0)$){cell\_y\_elements};
\end{scope}

\draw[dashed,gray] (hcell.north east) -- ([xshift=80mm]0,22);
\draw[dashed,gray] (hcell.south east) -- ([xshift=80mm]0,-22);

\end{tikzpicture}
\caption{Schematic of a SpaCal module showing the simulated volumes and the configuration-file keys (monospace) that control the dimensions of each element. \textbf{Top:} longitudinal section along the beam direction $z$. The absorber, embedded with longitudinal scintillating fibres (red), is a box of constant transverse size, and is sandwiched between the Interface and Readout volumes. Thin gaps with configurable material are inserted at every volume boundary (\texttt{gap\_size}, \texttt{abs\_interface\_extraGap}). The Interface and Readout volumes, together with the gaps, are trapezoidal: they match the absorber cross-section at their inner face and taper towards the outer faces of the module with the angle set by \texttt{slant\_angle\_deg}, so that modules can be placed at a stereo angle in the calorimeter without overlapping. The taper is exaggerated here for clarity. The two ends need not be symmetric: the lengths of the Interface and Readout volumes at the negative and positive $z$ ends are set independently through the four keys shown, so that the model can be matched to the actual dimensions of the module being constructed. When split into front and back longitudinal sections for shower-depth profiling, the absorber is divided by a configurable separator. \textbf{Bottom left:} face-on view of the absorber, segmented into a grid of readout cells. \textbf{Bottom right:} zoom on a single cell, in which the scintillating fibres are arranged on a regular lattice. Here round cross-section fibres are shown, but square fibres can also be used in the framework. The optional cladding (single or double, made of PMMA + fluorinated polymer) around each fibre is controlled by the \texttt{crystal\_inner\_cladding\_fraction} and \texttt{crystal\_outer\_cladding\_fraction} keys (not drawn at this scale).}
\label{fig:spacal_schematic}
\end{figure*}

A SpaCal module is constructed as a composite solid, featuring a rectangular central absorber box joined with trapezoidal interface and readout volumes at both ends (see Figure~\ref{fig:spacal_schematic}). This tapered shape accommodates angled stereo placements within the calorimeter without causing volume overlaps. The dense absorber matrix (e.g. tungsten, lead) is embedded with longitudinal scintillating fibres grouped into readout cells. Fibres and their corresponding holes can be configured as either square or cylindrical. To accurately simulate total internal reflection, the framework supports nested concentric structures, including double-cladded fibres featuring an inner PMMA layer and an outer fluorinated polymer layer. Boundary interactions are governed by the Geant4 UNIFIED model. Scintillator faces can be assigned a specific surface roughness parameter ($\sigma_{\alpha}$) to emulate mechanical depolishing, while reflectors like ESR foils are modelled as \emph{dielectric\_metal} interfaces using empirical complex refractive index data.

\begin{figure*}[pos=tbp]
\centering
\begin{tikzpicture}[
    font=\scriptsize, >=Latex,
    x=1mm, y=1mm,
    ann/.style={font=\scriptsize\ttfamily}
]
\def\moduleH{20}

\fill[cyan!20,draw=black,thin] (0,0) rectangle (5.5,\moduleH);

\foreach \i in {0,1,2,3,4}{
  \pgfmathsetmacro{\xs}{5.5 + \i*4.8}
  \pgfmathsetmacro{\xa}{\xs + 1.0}
  \pgfmathsetmacro{\xt}{\xa + 0.4}
  \pgfmathsetmacro{\xe}{\xt + 3.0}
  \fill[gray!55,draw=black,thin]   (\xs,0) rectangle (\xa,\moduleH); %
  \fill[white,draw=black,thin]     (\xa,0) rectangle (\xt,\moduleH); %
  \fill[orange!35,draw=black,thin] (\xt,0) rectangle (\xe,\moduleH); %
  \fill[white,draw=black,thin]     (\xe,0) rectangle (\xe+0.4,\moduleH); %
}

\fill[red!55,draw=black,thin] (29.5,0) rectangle (32,\moduleH);

\foreach \i in {0,1,2,3,4}{
  \pgfmathsetmacro{\xs}{32 + \i*4.8}
  \pgfmathsetmacro{\xa}{\xs + 1.0}
  \pgfmathsetmacro{\xt}{\xa + 0.4}
  \pgfmathsetmacro{\xe}{\xt + 3.0}
  \fill[gray!55,draw=black,thin]   (\xs,0) rectangle (\xa,\moduleH);
  \fill[white,draw=black,thin]     (\xa,0) rectangle (\xt,\moduleH);
  \fill[orange!35,draw=black,thin] (\xt,0) rectangle (\xe,\moduleH);
  \fill[white,draw=black,thin]     (\xe,0) rectangle (\xe+0.4,\moduleH);
}

\fill[cyan!20,draw=black,thin] (56,0) rectangle (61.5,\moduleH);

\foreach \y in {3,8,13,18}{
  \draw[red!75,line width=0.45pt] (0,\y) -- (29.5,\y);
}
\foreach \y in {3,8,13,18}{
  \draw[red!75,line width=0.45pt] (32,\y) -- (61.5,\y);
}

\draw[->,thick] (-2,-9) -- (63,-9) node[right]{$z$};

\node[ann,above=2pt] (shthick) at (70,-7) {shashlik\_window\_thickness};
\draw[->,thin,gray] (shthick.north) to[bend left=15] (60,0);

\draw[decorate,decoration={brace,amplitude=4pt,raise=2pt}]
   (5.5,\moduleH) -- (29.5,\moduleH)
   node[midway,above=8pt,ann]{shashlik\_tiles\_in\_front\_section};

\draw[decorate,decoration={brace,mirror,amplitude=4pt,raise=2pt}]
   (5.5,0) -- (56,0)
   node[midway,below=11pt,ann]{shashlik\_number\_of\_layers};

\node[ann,align=left] (seplab) at (60,\moduleH+10)
   {shashlik\_separationThickness};
\draw[->,thin,gray] (seplab.south) to[bend left=15] (30.75,\moduleH);

\node[ann,align=left] (wlslab) at (70,22)
   {shashlik\_wls\_fiber\_radius};
\draw[->,thin,gray] (wlslab.south) -- (61.6,13);

\draw[<->,thin] (0,-12) -- (61.5,-12);
\node[ann,below=1pt] at (30.75,-12.5) {shashlik\_module\_size\_z (auto-computed)};

\end{tikzpicture}

\vspace{6mm}

\begin{tikzpicture}[
    font=\scriptsize, >=Latex,
    x=1mm, y=1mm,
    ann/.style={font=\scriptsize\ttfamily}
]
\def\zH{32}

\fill[gray!55,draw=black,thin]   (0,0)    rectangle (7,\zH);
\fill[white,draw=black,thin]     (7,0)    rectangle (10.5,\zH);
\fill[orange!35,draw=black,thin] (10.5,0) rectangle (26.5,\zH);
\fill[white,draw=black,thin]     (26.5,0) rectangle (30,\zH);

\draw[red!75,line width=0.5pt] (0,16) -- (30,16);

\node[ann,align=left] (holelab) at (-15,26) {through-hole:\\air gap + cladding\\+ WLS core};
\draw[->,thin,gray] (holelab.south) to[bend right=15] (0,16);

\node[ann,anchor=south east] (pb_name) at (7,\zH+2) {Pb absorber};
\draw[->,thin,gray] (pb_name.south) -- (3.5,\zH+0.5);
\node[ann,anchor=south west] (tile_name) at (10.5,\zH+2) {Polystyrene tile};
\draw[->,thin,gray] (tile_name.south) -- (18.5,\zH+0.5);
\node[ann,anchor=south west] (tyvek_name) at (10.5,\zH+6) {Tyvek wrap};
\draw[->,thin,gray] (tyvek_name.west) -- (8.5,\zH+0.5);
\draw[->,thin,gray] (tyvek_name.east) -- (28.5,\zH+0.5);

\draw[<->,thin] (0,-3) -- (7,-3);
\node[ann,anchor=north east] at (7.5,-3.2) {shashlik\_absorber\_thickness};

\draw[<->,thin] (10.5,-7) -- (26.5,-7);
\node[ann,anchor=north] at (18.5,-7.2) {shashlik\_tile\_size\_z};

\node[ann] (tylab) at (22,-14) {shashlik\_tyvek\_thickness};
\draw[->,thin,gray] (tylab.north west) to[bend left=15] (8.5,-0);
\draw[->,thin,gray] (tylab.north east) to[bend right=15] (28.6,-0);

\node[font=\scriptsize] at (15,\zH+12) {Repeating element (zoom)};

\end{tikzpicture}

\caption{Schematic of a Shashlik module showing the simulated volumes and the configuration-file keys (monospace) that control the dimensions of each element. The readout and interface volumes are defined in the same way as in the SpaCal modules (see Fig. \ref{fig:spacal_schematic}), and are not drawn here. \textbf{Top:} longitudinal section along the beam direction $z$. The module is built from a stack of \texttt{shashlik\_number\_of\_layers} repeating elements, i.e. one lead absorber plate plus one Tyvek-wrapped polystyrene scintillator tile, sandwiched between front and back windows. When split into front and back longitudinal sections, the stack is divided by a configurable separator after \texttt{shashlik\_tiles\_in\_front\_section} layers. WLS fibres (red) run longitudinally through the entire stack, threading all the absorber plates and scintillator tiles. \textbf{Bottom:} zoom on a single repeating element showing the nested layer structure (Pb absorber, air gap, Tyvek wrap, polystyrene tile, Tyvek wrap, air gap) and a representative WLS through-hole modelled as a nested air-gap + cladding + active core.}
\label{fig:shashlik_schematic}
\end{figure*}
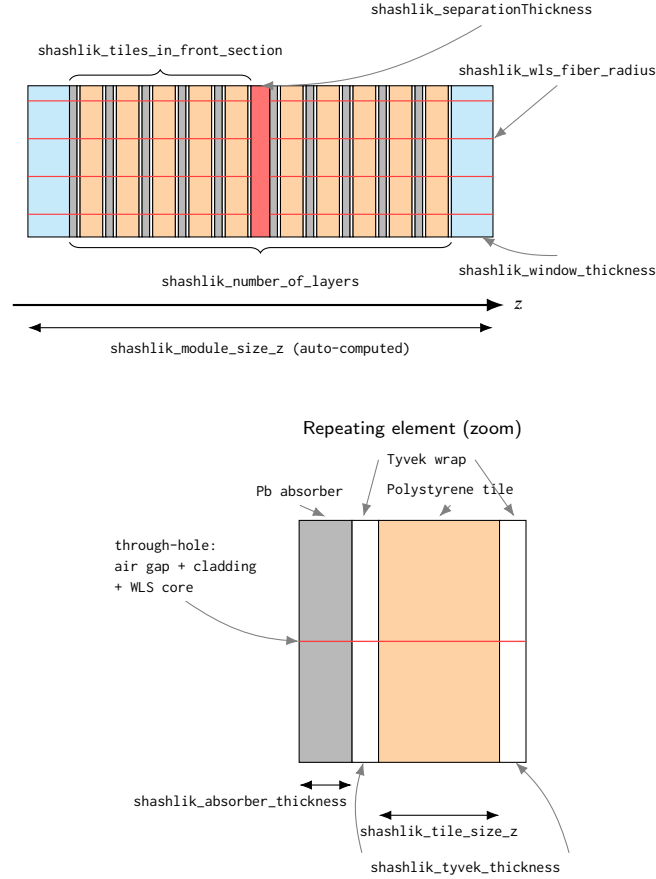

A Shashlik module is built hierarchically using a repeating element consisting of one lead absorber plate and one Tyvek-wrapped polystyrene scintillator layer, sandwiched between structural front and back endcaps. Within each scintillating layer, the polystyrene is segmented into an $N_x \times N_y$ array of discrete tiles housed within a container. Scintillation light is captured by WLS fibres running longitudinally along the entire layer stack. The framework automatically calculates the necessary through-holes across all materials, modelling each hole as a nested structure comprising an air gap, a polymeric cladding layer, and the active WLS core. The resulting longitudinal structure, together with the configuration-file keys that control each element, is shown in Figure~\ref{fig:shashlik_schematic}.

For both SpaCal and Shashlik technologies, modules can be split into front and back sections. These sections are optically isolated using configurable separators such as air gaps, aluminisation, or ESR foils.
For single-sided readout studies, a mirror is placed on the non-instrumented face to reflect light back towards the active photodetectors.

The LHCb ECAL spans about $6.3\times7.8$~m$^2$ and comprises $3312$ modules arranged on a $64\times52$ grid, with a central $4\times4$-module opening left to accommodate the beam plug \cite{CERN-LHCC-2000-036}. This overall geometry is common to the current and the upgraded calorimeter. The Upgrade~II ECAL will mix SpaCal and Shashlik technologies across the same grid. To simulate the entire detector, the framework scales up using a hierarchical configuration system. The global layout is dictated by a 2D map (provided as a ROOT histogram or text file) that assigns the exact spatial coordinates, module types, and stereo rotation angles across the calorimeter face.

The entire framework is steered through plain-text configuration files. No recompilation is required to switch between simulation modes or to study a new module. Listing~\ref{lst:cfg} reproduces a representative excerpt from the configuration file of a W-Polystyrene SpaCal module. The \texttt{simulationType} key selects one of the four supported modes (energy-deposition only, full Geant4 ray-tracing, hybrid simulation, optical calibration), making the same code base usable for both the production runs and the preparatory optical-calibration step.

\begin{figure}[pos=tbp]
\begin{lstlisting}[style=hmcconfig,label={lst:cfg},caption={Excerpt of a representative HybridMC configuration file (W-Polystyrene SpaCal module). The \texttt{simulationType} key switches between energy-deposition, full ray-tracing, hybrid, and optical-calibration modes without recompilation. Material codes (e.g. \texttt{8} for pure tungsten, \texttt{12} for polystyrene) are documented inline in the framework distribution.}]
# Simulation mode
#   0 = energy deposition only
#   1 = full Geant4 ray-tracing
#   2 = hybrid simulation
#   3 = optical calibration
simulationType = 2

# Absorber: 121.2 x 121.2 x 190 mm^3 tungsten matrix
absorber_size_x   = 121.2
absorber_size_y   = 121.2
absorber_size_z   = 190.
absorber_material = 8   # Pure tungsten, 19.1 g/cm^3

# One readout cell: 12 x 12 polystyrene fibres
cell_x_elements        = |12|
cell_y_elements        = |12|
cell_crystal_size_x    = |1.00|
cell_crystal_size_y    = |1.00|
cell_crystal_size_z    = |190.00|
cell_crystal_pitch_x   = |1.67|
cell_crystal_pitch_y   = |1.67|
cell_crystal_material  = |12|   # Polystyrene
cell_crystal_cladding  = |1|    # single clad 

# Optical handling
switchOnScintillation  = 1
propagateScintillation = 1
switchOnCerenkov       = 1
propagateCerenkov      = 1
\end{lstlisting}
\end{figure}

\subsection{Optical calibration}\label{sec:opticalcalibration}

\begin{figure*}[pos=htbp]
\centering
\begin{tikzpicture}[
    font=\scriptsize, >=Latex, x=1mm, y=1mm,
    ann/.style={font=\scriptsize\ttfamily}
]
\def\moduleH{12}
\def\winW{4}
\def\frontLen{30}
\def\sepW{2}
\def\backLen{44}
\pgfmathsetmacro{\bsStart}{\winW+\frontLen+\sepW}
\pgfmathsetmacro{\bbEnd}{\bsStart+\backLen}

\fill[green!22,draw=black,thin] (0,0) rectangle (\winW,\moduleH);
\node[ann,below=1pt,align=center] at (\winW/2,-1.5) {readout$^-$};

\fill[yellow!22,draw=black,thin] (\winW,0) rectangle (\winW+\frontLen,\moduleH);
\node[ann,above=1pt] at (\winW+\frontLen/2,\moduleH) {fibre\ front};

\fill[red!50,draw=black,thin] (\winW+\frontLen,0) rectangle (\winW+\frontLen+\sepW,\moduleH);

\fill[yellow!32,draw=black,thin] (\bsStart,0) rectangle (\bsStart+\backLen,\moduleH);
\node[ann,above=1pt] at (\bsStart+\backLen/2,\moduleH) {fibre\ back};

\fill[green!22,draw=black,thin] (\bbEnd,0) rectangle (\bbEnd+\winW,\moduleH);
\node[ann,below=1pt,align=center] at (\bbEnd+\winW/2,-1.5) {readout$^+$};

\foreach \x in {8, 14, 20, 26, 32}{
  \foreach \y in {3, 6, 9}{
    \fill[gray!70] (\x,\y) circle (0.3);
  }
}
\foreach \x in {42, 48, 54, 60, 66, 72, 78}{
  \foreach \y in {3, 6, 9}{
    \fill[gray!70] (\x,\y) circle (0.3);
  }
}

\fill[red] (20,6) circle (0.6);
\foreach \a in {0,22.5,45,67.5,90,112.5,135,157.5,180,202.5,225,247.5,270,292.5,315,337.5}{
  \draw[->,red!75,thin,line width=0.35pt] (20,6) -- ++({3*cos(\a)},{3*sin(\a)});
}

\node[ann,below right,align=left] (ptlab) at (22,-7) {grid node:\\monochromatic\\isotropic emission};
\draw[->,gray,thin] (ptlab.north) to[bend right=15] (20.5,4.8);

\node[ann,align=left] (gridlab) at (60,-5) {grid: ($x$,\,$y$,\,$z$,\,$E$,\,$\hat{p}_z$)};
\draw[->,gray,thin] (gridlab.north) to[bend left=15] (54,3);

\draw[->,thick] (-1,-2) -- (\bbEnd+\winW+1,-2) node[right]{$z$};

\def\pdfH{16}
\def\pdfW{36}

\begin{scope}[xshift=98mm,yshift=14mm]
  \draw[->,thin] (0,0) -- (\pdfW+2,0) node[right,font=\tiny]{$t_\mathrm{arr}$};
  \draw[->,thin] (0,0) -- (0,\pdfH+2) node[above,font=\tiny]{PDF};
  \foreach \x/\h in {1.5/13.5, 3.5/10.5, 5.5/8, 7.5/6, 9.5/4.5, 11.5/3.6, 13.5/2.8, 15.5/2.2, 17.5/1.7, 19.5/1.3, 21.5/1.0, 23.5/0.8, 25.5/0.6, 27.5/0.5, 29.5/0.4}{
    \fill[blue!55] (\x,0) rectangle ({\x+1.6},\h);
  }
  \node[font=\scriptsize,anchor=south west] at (1,\pdfH+0.5) {Front readout};
\end{scope}

\begin{scope}[xshift=98mm,yshift=-12mm]
  \draw[->,thin] (0,0) -- (\pdfW+2,0) node[right,font=\tiny]{$t_\mathrm{arr}$};
  \draw[->,thin] (0,0) -- (0,\pdfH+2) node[above,font=\tiny]{PDF};
  \foreach \x/\h in {3.5/1.5, 5.5/3.5, 7.5/6.5, 9.5/8, 11.5/6.5, 13.5/4.8, 15.5/3.6, 17.5/2.7, 19.5/2.1, 21.5/1.6, 23.5/1.3, 25.5/1.0, 27.5/0.8, 29.5/0.6, 31.5/0.5}{
    \fill[green!40!black,fill opacity=0.55] (\x,0) rectangle ({\x+1.6},\h);
  }
  \node[font=\scriptsize,anchor=south west] at (1,\pdfH+0.5) {Back readout};
\end{scope}

\draw[->,gray!70!black,thin] (\winW+0.5,\moduleH-1) to[bend left=8] (98,28);
\draw[->,gray!70!black,thin] (\bbEnd-0.5,\moduleH-1) to[bend left=4] (98,2);

\end{tikzpicture}
\caption{Schematic of the optical-calibration procedure, shown for the case of a longitudinally segmented SpaCal module. The drawing does \emph{not} depict the entire multi-fibre module: thanks to the in-section symmetry of the SpaCal geometry (see text), a single representative scintillating fibre per section is sufficient to characterise the optical transport of all fibres in that section, and the figure therefore represents one such fibre (front and back longitudinal segments, in yellow) read out by photo-detector volumes (green) at both ends. \textbf{Left:} the fibre is sampled on a multidimensional grid of emission points spanning the spatial coordinates $(x,y,z)$, the monochromatic emission energy $E$, and the longitudinal momentum component $\hat{p}_z$. At every grid node (grey dots), $\sim$$2\!\times\!10^6$ photons are generated isotropically (red rays) and tracked by Geant4 through the full geometry until they are absorbed at the front or back readout boundary. \textbf{Right:} the resulting per-node histograms of arrival time at the front and back readout interfaces form a pair of probability density functions (\textsc{pdf}s). Their absolute normalisation is the photon extraction probability and their shape encodes the time-of-arrival distribution used by the parametrised optical-transport step.}\label{fig:opticali_example}
\end{figure*}
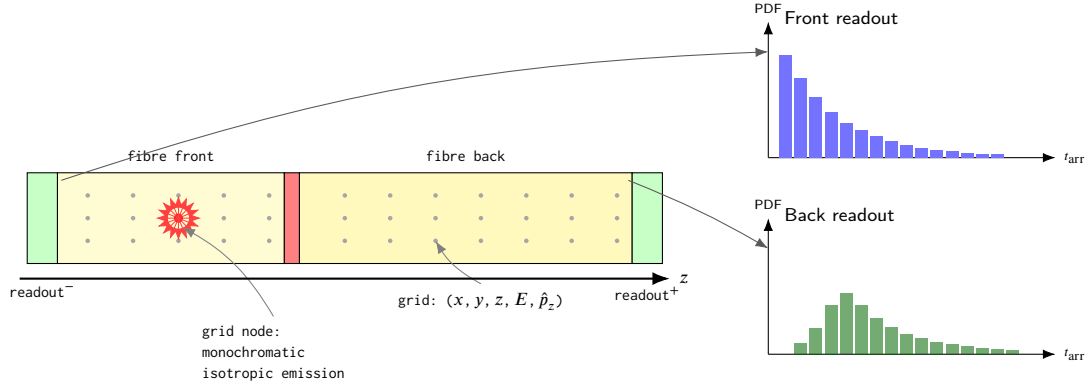

\begin{figure*}[pos=tbp]
\centering
\begin{tikzpicture}[
    font=\scriptsize, x=1mm, y=1mm,
    ann/.style={font=\scriptsize\ttfamily},
    ph/.style={
        draw, dashed, rounded corners,
        fill=gray!6, align=center, inner sep=1.5mm,
        execute at begin node=\sffamily
    }
]
\def\moduleH{12}
\def\winW{3}
\def\frontLen{56}
\def\sepW{2}
\def\backLen{70}
\pgfmathsetmacro{\bsStart}{\winW+\frontLen+\sepW}
\pgfmathsetmacro{\bbEnd}{\bsStart+\backLen}

\fill[green!22,draw=black,thin] (0,0) rectangle (\winW,\moduleH);
\fill[yellow!22,draw=black,thin] (\winW,0) rectangle (\winW+\frontLen,\moduleH);
\fill[red!50,draw=black,thin] (\winW+\frontLen,0) rectangle (\winW+\frontLen+\sepW,\moduleH);
\fill[yellow!32,draw=black,thin] (\bsStart,0) rectangle (\bsStart+\backLen,\moduleH);
\fill[green!22,draw=black,thin] (\bbEnd,0) rectangle (\bbEnd+\winW,\moduleH);

\node[ann,below=1pt,align=center] at (\winW/2,0)        {readout$^-$};
\node[ann,below=1pt,align=center] at (\bbEnd+\winW/2,0) {readout$^+$};
\node[ann,above=1pt] at (\winW+\frontLen/2,\moduleH + 3) {fibre\ front};
\node[ann,above=1pt] at (\bsStart+\backLen/2,\moduleH + 3) {fibre\ back};

\foreach \i/\x in {1/10, 2/20, 3/30, 4/40, 5/50}{
  \fill[blue!75!black] (\x,\moduleH/2) circle (0.8);
  \node[font=\tiny, blue!50!black, above=2pt] at (\x,\moduleH) {F\i};
}
\foreach \i/\x in {1/70, 2/82, 3/94, 4/106, 5/118}{
  \fill[red!75!black] (\x,\moduleH/2) circle (0.8);
  \node[font=\tiny, red!50!black, above=2pt] at (\x,\moduleH) {B\i};
}

\draw[->,thick] (-1,-7) -- (\bbEnd+\winW+1,-7) node[right]{$z$};

\end{tikzpicture}
\includegraphics[width=\textwidth]{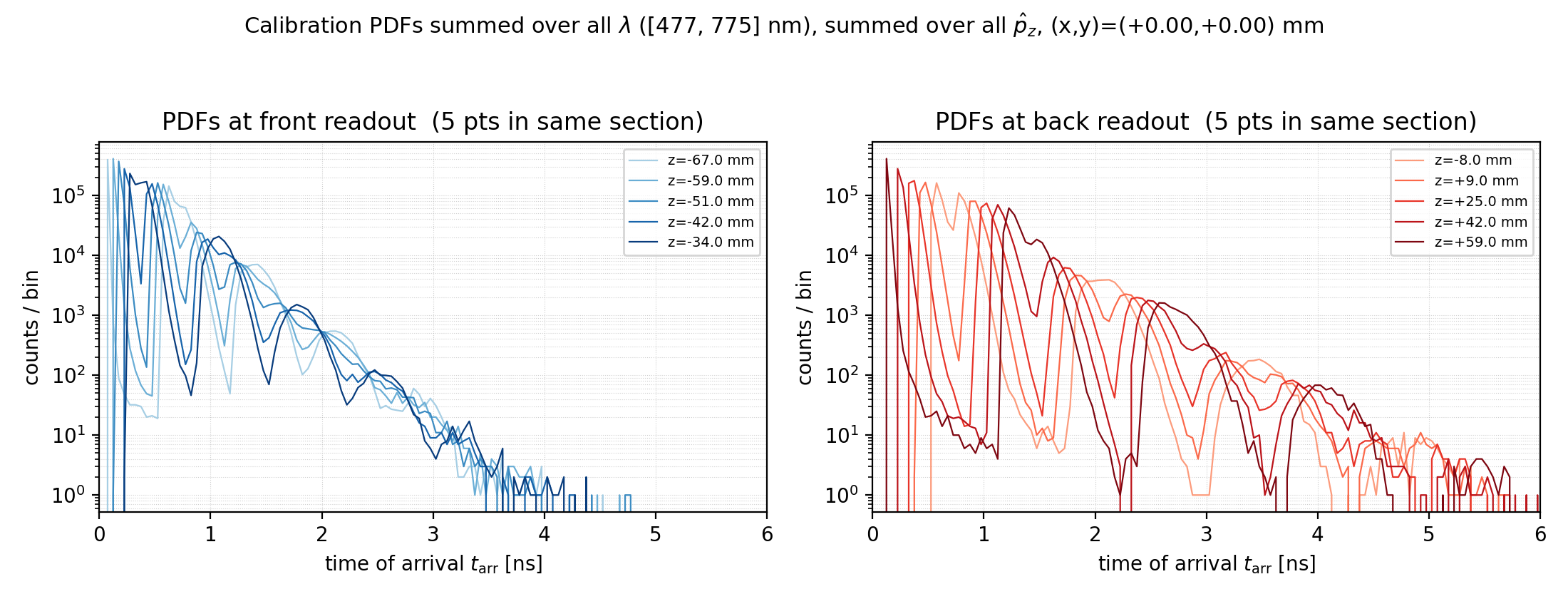}

\caption{Calibration PDFs extracted from a W-GAGG representative SpaCal configuration at five emission points placed along the front longitudinal section (F1\,\textendash\,F5, blue) and five along the back section (B1\,\textendash\,B5, red). The locator at the top shows the corresponding positions inside the calibrated fibre. The module in this example is 151 mm long in the z direction, (50 mm front section, 1 mm separation and 100 mm back section), with front/back separation located at z = -25 mm. \textbf{Left:} time-of-arrival PDFs recorded at the front readout interface, all five points overlaid. \textbf{Right:} same for the back readout interface.}
\label{fig:pdf_examples}
\end{figure*}

The calibration procedure consists of a full ray-tracing simulation executed within the Geant4 toolkit. Monochromatic optical photons are generated across a finely segmented multidimensional grid encompassing the spatial coordinates $(x,y,z)$, emission energies, and the longitudinal emission angle (parametrised as the momentum z-component). An example, for the case of a longitudinally segmented SpaCal module, is shown in Figure \ref{fig:opticali_example}. Typical configurations employ 1 mm spatial steps along the longitudinal axis and 0.1 eV energy intervals, thoroughly spanning the emission spectrum of the active scintillating medium. To suppress statistical fluctuations that could artificially degrade the simulated energy resolution, approximately $2\times10^{6}$ photons are generated isotropically at each grid point. During this phase, Geant4 evaluates all complex optical phenomena, including Fresnel refractions, total internal reflection, bulk absorption, and stochastic scattering caused by scintillator surface roughness. The propagation of each photon concludes when it crosses into a designated readout boundary volume, where it is absorbed and logged as a detected signal.
For every node in the multidimensional grid, the procedure outputs two 1-D histograms storing the time of arrival of the photons collected on the front and back readout interfaces. These histograms encode the two main features that are needed from the transport of optical photons in the specific technology under study. The photon extraction efficiency, defined as the absolute probability that a photon generated at a specific spatial coordinate, energy, and direction will reach a photodetector face, can be derived from the ratio of successfully recorded entries in either histogram to the total number of generated photons. The precise arrival times of the extracted photons can also be derived from the histograms, considering that they represent the probability density function of the time transport. Examples of such PDFs extracted from a representative SpaCal configuration are reported in Figure~\ref{fig:pdf_examples} for five emission points distributed along the front longitudinal section and five along the back section.

A comprehensive optical calibration needs to be performed for each module under study, and therefore in the case of full calorimeters made up of multiple module technologies, multiple calibrations need to be produced. Given the substantial computational requirements of this full ray-tracing procedure, the calibration campaign is heavily parallelised across distributed computing clusters. Upon completion of the parallel jobs, the outputs are aggregated, and the final PDF histograms are serialised into a custom binary format.

The dimensionality of the calibration grid depends on the technology. In a SpaCal module, the absorber matrix is embedded with an array of identical scintillating fibres, and the optical-transport history of a photon emitted inside one of these fibres is determined entirely by its position within the fibre, its energy, and its emission direction. The surrounding fibres and the absorber walls act only as boundaries. As a consequence, all fibres of a given longitudinal section can be described by the optical calibration of a single representative fibre, and the calibration grid is only required to span the interior of one fibre per section. This is the configuration depicted in Figure~\ref{fig:opticali_example}.

The Shashlik technology, by contrast, does not enjoy this in-section symmetry: each scintillator tile occupies a distinct position relative to the WLS-fibre lattice and the module boundaries, so the optical-transport history of photons emitted in different tiles is, in general, not equivalent. The calibration must therefore be performed independently for every tile of each longitudinal section. This both makes the resulting calibration table consistently larger than its SpaCal counterpart and, to keep the overall table size manageable, forces a sparser spatial sampling of the emission points within each tile than the dense grid used for the single representative SpaCal fibre. In practice, each Shashlik calibration file is of the order of one~GB, while the SpaCal files are in the range of 10-100~MB. This fits comfortably in the Random Access Memory (RAM) of a typical computing node, and the run-time overhead remains negligible even when several module types are loaded simultaneously to simulate the full ECAL.

\subsection{Energy deposition}\label{sec:edep}

The event-by-event simulation pipeline commences with the energy deposition phase. During this stage, the Geant4 simulation is exclusively dedicated to performing a full Monte Carlo tracking of the incident primary particles and the subsequent development of the electromagnetic shower within the calorimeter geometry, while the generation and propagation of scintillation and Cherenkov photons are strictly deactivated.

The simulation records a highly granular profile of the shower. For every interaction occurring within the simulation geometry, the algorithm extracts and records the energy deposited, coupled with the precise 3D global and local spatial coordinates, the local timestamp of the interaction, the specific material involved, and the respective module and track identifiers. This comprehensive dataset is exported as a 3D energy deposition map, stored as a ROOT TTree, which serves as the input for the subsequent optical parametrisation.

\begin{figure*}[pos=tbp]
\centering

\includegraphics[width=0.32\textwidth]{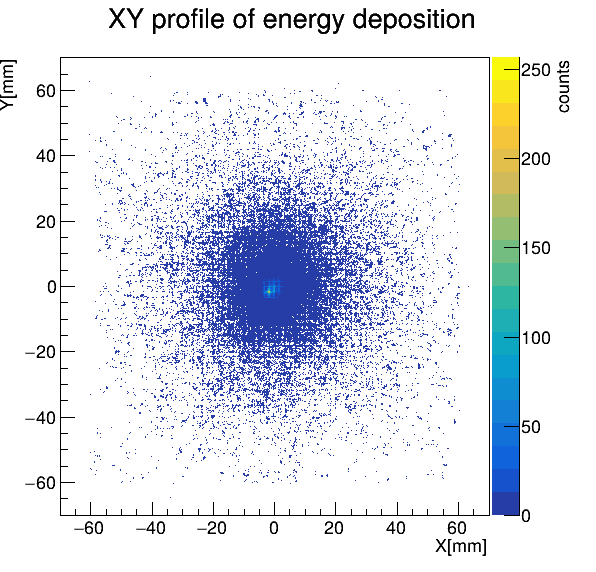}
\includegraphics[width=0.32\textwidth]{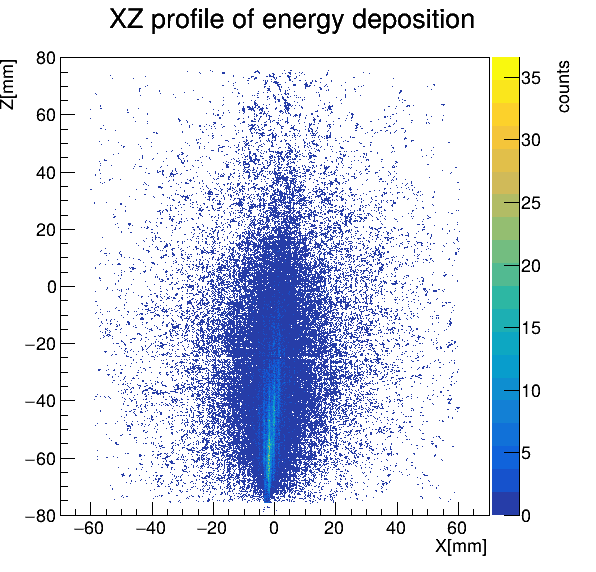}
\includegraphics[width=0.32\textwidth]{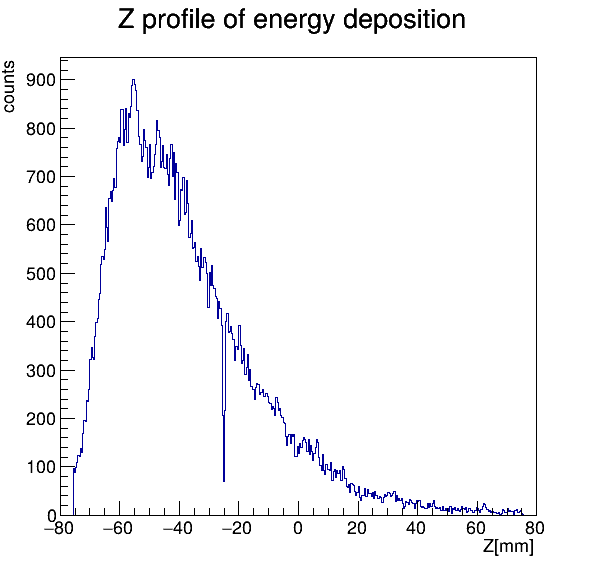}

\caption{Energy-deposition maps produced by the energy-deposition step for 100 $1$~GeV$/c$ electrons at the centre of a SpaCal module, with 3+3 degrees beam inclination. \textbf{Left:} transverse $(x,y)$ projection of the deposited energy. \textbf{Middle:} longitudinal $(x,z)$ projection. \textbf{Right:} 1-D profile of energy deposited along the $z$ axis.}
\label{fig:edep_example_spacal}
\end{figure*}

\begin{figure*}[pos=tbp]
\centering

\includegraphics[width=0.32\textwidth]{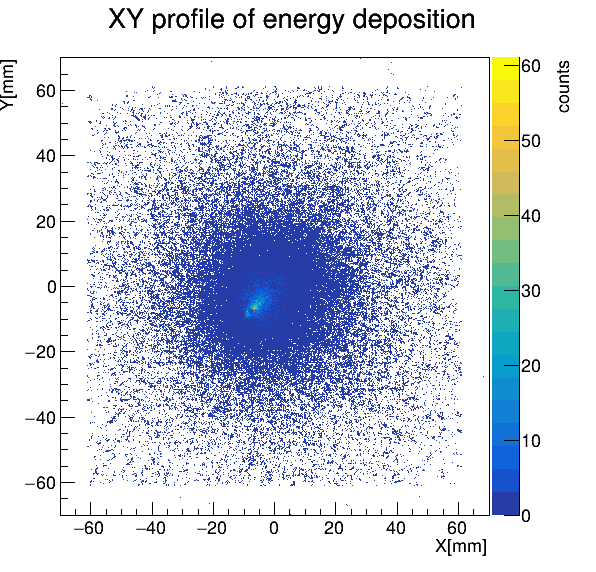}
\includegraphics[width=0.32\textwidth]{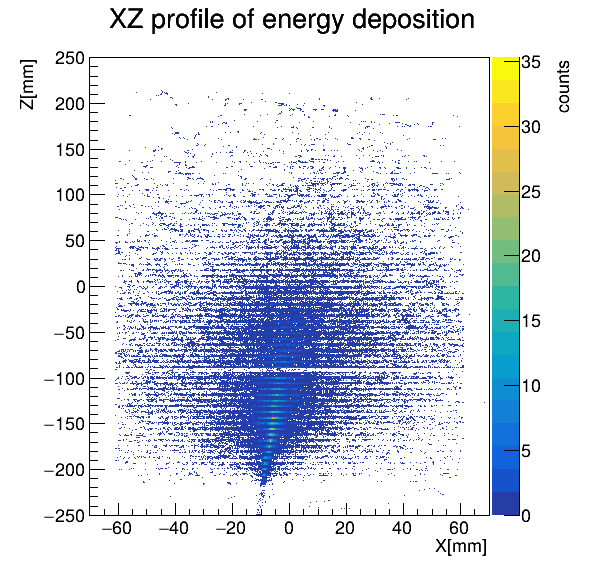}
\includegraphics[width=0.32\textwidth]{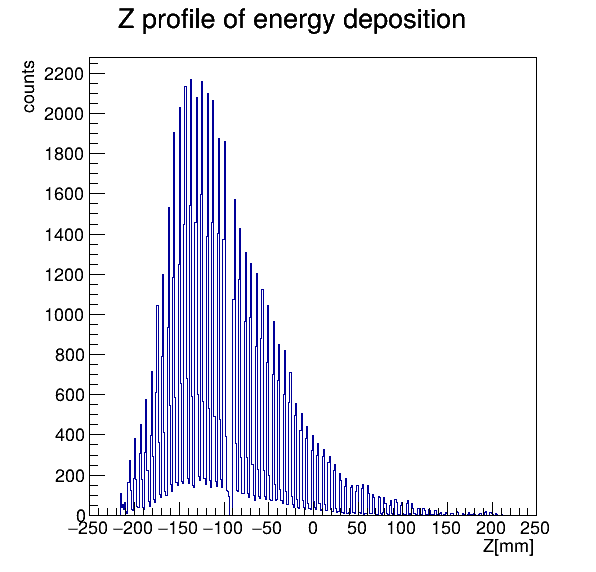}

\caption{Energy-deposition maps produced by the energy-deposition step for 100 $1$~GeV$/c$ electrons at the centre of a Shashlik module, with 3+3 degrees beam inclination. \textbf{Left:} transverse $(x,y)$ projection of the deposited energy. \textbf{Middle:} longitudinal $(x,z)$ projection. \textbf{Right:} 1-D profile of energy deposited along the $z$ axis.}
\label{fig:edep_example_shashlik}
\end{figure*}

As a representative example of the output produced by the energy-deposition step, Figures~\ref{fig:edep_example_spacal} and~\ref{fig:edep_example_shashlik} show the 3-D energy-deposition maps recorded by the framework for a SpaCal and a Shashlik module, respectively. In both cases the maps accumulate the showers of $100$ electrons of $1$~GeV$/c$, shot consecutively at the centre of the module with a direction that forms a $3+3$ degrees stereo angle with respect to the main axis of the module. For each module type, the transverse $(x,y)$ projection, the longitudinal $(x,z)$ projection, and the 1-D energy profile along the beam direction are displayed.

A critical feature of this phase is the specialised handling of Cherenkov radiation. Because Cherenkov photons are emitted promptly and possess strong directionality, they significantly influence the leading edge of the photodetector pulse and, consequently, the ultimate time resolution. To accurately preserve this timing information without incurring the penalty of full ray-tracing, Cherenkov photons are generated by Geant4 according to standard kinematics, but are instantaneously terminated at birth. The simulation logs their exact birth coordinates, emission energies, momentum directions, and generation timestamps into a dedicated data structure, later used for parametrised transport.

\subsection{Parametrised optical transport}\label{sec:pot}

Following the generation of the highly granular energy deposition map, the framework executes the algorithm for parametrised transport of optical photons, which constitutes the principal acceleration mechanism of the HybridMC paradigm. Upon initialisation, the algorithm loads the binary calibration files (.data) into RAM.

As explained above, the main output of the energy deposition step of the framework is composed of two different lists, namely the energy deposition maps and the list of Cherenkov photons generated. The parametrised optical transport treats these two lists with different algorithms, described below.

\subsubsection{Transport of scintillation photons}

The routine iterates event-by-event over every discrete energy deposition step recorded in the output of the energy deposition step. These energy depositions would normally create scintillation photons in the active materials. For each interaction, therefore, the algorithm computes an expected number of scintillation photons produced. This calculation is performed taking into account the deposited energy, the active material's empirical light yield, and its Birks' quenching coefficient~\cite{Birks1951}. For the polystyrene-based scintillators a Birks coefficient of $0.126$~mm/MeV is used~\cite{Badhwar1967}. For GAGG the quenching is neglected, as it is usually done for high-$Z$ inorganic scintillators. Although recent measurements found a non-zero values for some GAGG samples, e.g. $k_B=0.075$~mm/MeV~\cite{GAGG_Birks}, this effect is confined to the hard X-ray and soft gamma-ray regime and is negligible over the energy range reconstructed by the LHCb ECAL. For each generated photon candidate, the emission wavelength is sampled from the material's specific emission spectrum, and a binary search is performed to locate the nearest spatial $(x,y,z)$ and energy bin within the multidimensional calibration grid. A Monte Carlo roll is subsequently evaluated against the retrieved extraction efficiency for both the front and back readout interfaces. This probabilistic evaluation determines whether the photon successfully navigates the calorimeter geometry to reach a photodetector, implicitly accounting for all optical losses due to bulk absorption, surface roughness, and imperfect total internal reflection.

If a photon is deemed detected, its arrival timestamp at the photo-detector is synthesised as the linear sum of three independent temporal components:
\begin{equation}\label{eq:tarrival}
T_{\rm arrival} = T_{\rm dep} + T_{\rm gen} + T_{\rm prop},
\end{equation}
where $T_{\rm dep}$ is the global Geant4 time of the initial ionising interaction, $T_{\rm gen}$ is the stochastic scintillation emission delay sampled from the active material's multi-exponential decay profile, and $T_{\rm prop}$ is the optical transport time through the module geometry, drawn from the calibration probability density function described in Section~\ref{sec:framework}.

\subsubsection{Transport of Cherenkov photons}
\begin{figure*}[pos=tbp]
\centering
\begin{tikzpicture}[
    font=\small, >=Latex,
    x=1mm, y=1mm,
    box/.style={
        rectangle, rounded corners=2pt,
        draw, line width=0.6pt,
        minimum height=12mm, minimum width=27mm,
        align=center, inner sep=2pt
    },
    cal/.style    ={box, fill=blue!10,   draw=blue!60!black},
    inp/.style    ={box, fill=orange!18, draw=orange!70!black},
    sci/.style    ={box, fill=green!15,  draw=green!55!black},
    cer/.style    ={box, fill=cyan!18,   draw=cyan!60!black},
    tarr/.style   ={box, fill=red!10,    draw=red!60!black},
    detlist/.style={box, fill=red!18,    draw=red!60!black,
                    minimum width=24mm,  minimum height=24mm},
    flow/.style   ={->, thick, shorten >=1pt, shorten <=1pt},
    pdfarr/.style ={->, thin, dashed, blue!55!black,
                    shorten >=1pt, shorten <=1pt}
]

\node[inp]                                 (edep) at ( -55, 18) {Energy deposition\\map\\[1pt]\scriptsize from Geant4};
\node[sci, right=4mm of edep]              (nph)               {Scintillation photons\\per $\Delta E$\\[1pt]\scriptsize $N_\mathrm{ph}\propto$ LY $\cdot$ Birks};
\node[sci, right=4mm of nph]               (sciL)              {Sample $\lambda$;\\look up $\varepsilon_\mathrm{front},\varepsilon_\mathrm{back}$\\[1pt]\scriptsize indexed by $(x,y,z,E)$};
\node[tarr,right=4mm of sciL]              (sciT)              {$T_\mathrm{arr}=T_\mathrm{dep}{+}T_\mathrm{gen}{+}T_\mathrm{prop}$\\[1pt]\scriptsize $T_\mathrm{gen}$: multi-exp.\ decay\\$T_\mathrm{prop}$: PDF};

\node[inp]                                 (clist) at ( -55,-18) {Cherenkov\\photon list\\[1pt]\scriptsize from Geant4 (killed at birth)};
\node[cer, right=4mm of clist]             (ckin)               {Use Geant4\\kinematics directly\\[1pt]\scriptsize $(x,y,z,E,\hat p_z)$};
\node[cer, right=4mm of ckin]              (cerL)               {Look up\\$\varepsilon_\mathrm{front},\varepsilon_\mathrm{back}$\\[1pt]\scriptsize indexed by $(x,y,z,E,\hat p_z)$};
\node[tarr,right=4mm of cerL]              (cerT)               {$T_\mathrm{arr}=T_\mathrm{dep}{+}T_\mathrm{prop}$\\[1pt]\scriptsize (Cherenkov is prompt)};

\node[cal] (cal) at (sciL |- 0,0) {Calibration PDFs\\[1pt]\scriptsize \texttt{.data} loaded in RAM};

\node[detlist, right=10mm of sciT.east, yshift=-18mm] (det)
   {Detected\\photon list\\[1pt]\scriptsize $(t,\,x,\,y,\,\lambda)$\\$\rightarrow$ digitisation};

\draw[flow] (edep)  -- (nph);
\draw[flow] (nph)   -- (sciL);
\draw[flow] (sciL)  -- (sciT);
\draw[flow] (clist) -- (ckin);
\draw[flow] (ckin)  -- (cerL);
\draw[flow] (cerL)  -- (cerT);

\draw[flow] (sciT.east) -- ++(4mm,0) |- (det.north west);
\draw[flow] (cerT.east) -- ++(4mm,0) |- (det.south west);

\draw[pdfarr] (cal.north) -- (sciL.south);
\draw[pdfarr] (cal.south) -- (cerL.north);

\end{tikzpicture}
\caption{Schematic of the parametrised optical-transport step. The same calibration PDFs (top centre, loaded once into RAM) feed two parallel branches. \textbf{Top:} scintillation photons are generated on the fly from the Geant4 energy-deposition map. For every infinitesimal deposit the routine evaluates the expected number of scintillation photons (light yield times Birks' quenching factor), samples the emission wavelength from the material's spectrum, looks up the front and back extraction efficiencies in the calibration grid indexed by $(x,y,z,E)$, and rolls a Monte Carlo against them. Detected photons receive an arrival time built as $T_\mathrm{dep}+T_\mathrm{gen}+T_\mathrm{prop}$, where $T_\mathrm{gen}$ is sampled from the active material's multi-exponential decay and $T_\mathrm{prop}$ from the PDF. \textbf{Bottom:} Cherenkov photons are taken from the list pre-computed by Geant4 (which generates them but kills them at birth). Their deterministic kinematics, including the longitudinal momentum component $\hat p_z$, are used to look up the extraction efficiency in the calibration grid (now indexed by $(x,y,z,E,\hat p_z)$). Because Cherenkov emission is prompt, only $T_\mathrm{dep}+T_\mathrm{prop}$ contributes to the arrival time. The detected photons from both branches are merged into a single list that is forwarded to the digitisation and signal-formation step.}\label{fig:pot_scheme}
\end{figure*}
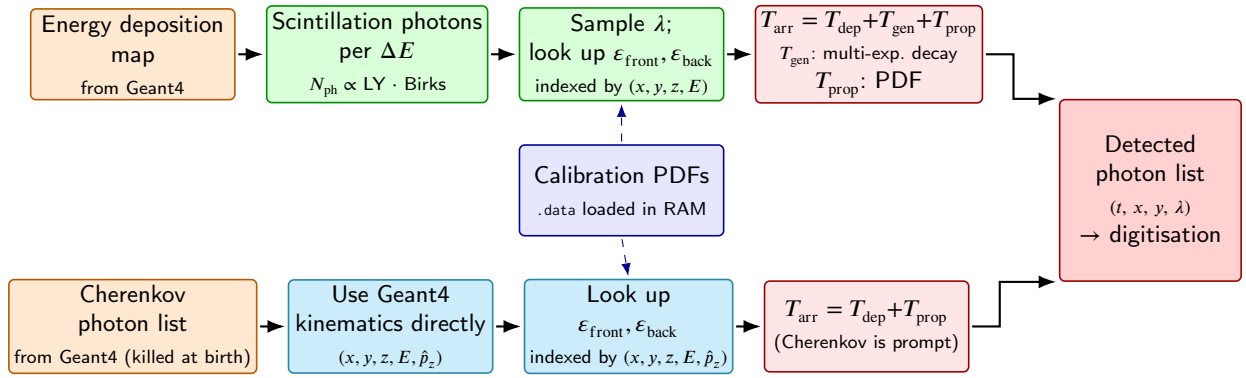

A distinct processing loop is dedicated to the Cherenkov photons previously generated and killed at birth by Geant4. Because these photons possess deterministic birth coordinates, energies, and momentum vectors, the algorithm bypasses the stochastic emission sampling phase. Their extraction probability is evaluated using also the momentum z-component bins of the calibration grid. The $T_{gen}$ component is set to $0$, to preserve the prompt emission nature of Cherenkov radiation. The full processing chain, with the two parallel scintillation and Cherenkov branches sharing the same calibration table, is illustrated in Figure~\ref{fig:pot_scheme}.

The output dataset of the Parametrised optical transport step of the framework, therefore, catalogues all successfully detected optical photons, their specific photodetector cell assignments, and their precise temporal and energetic kinematics, effectively preserving the high-fidelity optical physics of a full ray-tracing simulation while accelerating the transport process.

\subsection{Digitisation and signal formation}\label{sec:digitisation}

Following the parametrised optical transport, the framework executes the algorithm to transform the discrete ensemble of detected photon timestamps into waveforms.

The digitisation routine aggregates the simulated optical photons by event, module, and individual readout cell. Light losses occurring between the active scintillator volume and the photo-detector entrance window, which are not included in the optical calibration of the module, are then taken into account. In the simplest configuration, a single multiplicative loss factor is applied uniformly to the incident photon flux: this scalar absorbs the average effect of all such mechanisms, for instance the average transmission of an external light guide or the optical-grease coupling between the calibrated module and the photo-detector. For module designs where these losses are spatially non-uniform, the framework supports a configurable spatial-efficiency filter, which assigns a different detection probability to each point of the photo-detector entrance window and is therefore able to capture position-dependent effects such as the non-uniform light-collection efficiency of a hollow light guide. Subsequently, a Quantum Efficiency (QE) filter evaluates each surviving photon. The photon's wavelength is cross-referenced against the empirical QE curve of the specific photodetector model. A stochastic Monte Carlo roll determines the generation of a photoelectron.

For each successfully generated photoelectron, the algorithm synthesises the electronic response. A random temporal jitter is introduced to the photon's arrival time, sampled from the specific detector's Single-Photon Time Resolution (SPTR) distribution.
Then, for each photoelectron a Single-Photon Electron (SPE) pulse is produced. The SPE pulses are summed and the resulting waveform is sampled. To emulate a realistic data acquisition system, such as a DRS4 digitiser~\cite{Ritt2008DRS4}, this array is typically configured with 1024 sampling bins spanning a 204.8~ns temporal gate, yielding a highly granular 5~GS/s sampling rate ($200$~ps per sample). The number of bins, the gate width, the sampling rate, the baseline and the electronic-noise level are all set through the configuration file, so that the response of a specific readout system can be reproduced without recompiling the framework.
Gaussian noise parameters (e.g., 0.24 mV RMS noise or 1.7 ADC channels) are superimposed onto the final digitised waveform.

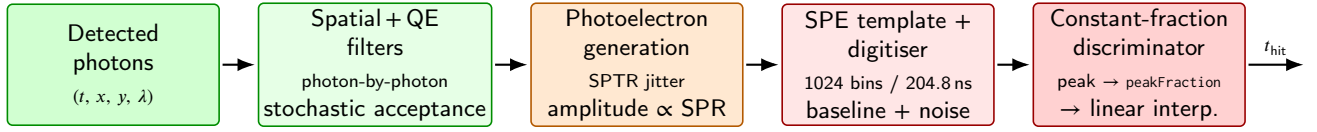
\begin{figure*}[pos=tbp]
\centering
\begin{tikzpicture}[
    font=\small, >=Latex,
    node distance=3mm and 5mm,
    box/.style={
        rectangle, rounded corners=2pt,
        draw, line width=0.6pt,
        minimum height=16mm, minimum width=28mm,
        align=center, inner sep=2pt
    },
    inp/.style    ={box, fill=green!18, draw=green!55!black},
    flt/.style    ={box, fill=green!10, draw=green!55!black},
    pmt/.style    ={box, fill=orange!18,draw=orange!70!black},
    dig/.style    ={box, fill=red!10,   draw=red!60!black},
    cfd/.style    ={box, fill=red!18,   draw=red!60!black},
    flow/.style   ={->, thick, shorten >=1pt, shorten <=1pt},
    icon/.style   ={font=\scriptsize, gray!60!black}
]

\node[inp]                  (in)  {Detected\\photons\\[1pt]\scriptsize $(t,\,x,\,y,\,\lambda)$};
\node[flt, right=of in]     (flt) {Spatial\,$+$\,QE\\filters\\[1pt]\scriptsize photon-by-photon\\stochastic acceptance};
\node[pmt, right=of flt]    (pe)  {Photoelectron\\generation\\[1pt]\scriptsize SPTR jitter \\amplitude $\propto$ SPR};
\node[dig, right=of pe]     (dig) {SPE template $+$\\digitiser\\[1pt]\scriptsize 1024 bins / 204.8\,ns\\baseline $+$ noise};
\node[cfd, right=of dig]    (cfd) {Constant-fraction\\discriminator\\[1pt]\scriptsize peak $\to$ \texttt{peakFraction}\\$\to$ linear interp.};

\draw[flow] (in)  -- (flt);
\draw[flow] (flt) -- (pe);
\draw[flow] (pe)  -- (dig);
\draw[flow] (dig) -- (cfd);

\draw[flow] (cfd.east) -- node[above, font=\scriptsize]{$t_\text{hit}$} ++(8mm,0);

\end{tikzpicture}
\caption{Schematic of the pulse-formation pipeline applied to every readout cell. Detected photons (with timestamp, position, and wavelength) are first filtered by configurable spatial-efficiency and quantum-efficiency curves of the photo-detector. Each surviving photon generates a single photoelectron whose arrival time is randomly displaced by the detector's Single-Photon Time Resolution (SPTR), and whose amplitude is drawn from the Single-Photon Electron response distribution. A photoelectron-response template is then superimposed onto a $1024$-bin discrete waveform array (e.g.~$5$~GS/s, $204.8$~ns gate), and Gaussian electronic noise is added bin-by-bin on top of a configurable baseline. The resulting waveform is finally processed by a Constant-Fraction Discrimination (CFD) algorithm: it estimates the baseline statistics from the pre-pulse samples, localises the signal peak above a signal-to-noise threshold, sets the trigger level at \texttt{peakFraction} of the peak amplitude, and returns the per-cell hit time $t_\text{hit}$ by linearly interpolating between the two samples bracketing the threshold crossing.}\label{fig:pulse_formation}
\end{figure*}

The final stage of the signal processing pipeline uses a Constant Fraction Discrimination (CFD) algorithm to extract high-precision timestamps from the synthesised macroscopic waveforms, emulating the trigger and timing electronics of the calorimeter. The algorithm estimates the baseline mean and noise from the initial samples. It subsequently identifies the signal peak (accounting for the configurable pulse polarity) and verifies that the maximum amplitude exceeds a pre-defined signal-to-noise threshold. A trigger level is established at a user-configurable fraction of the peak amplitude. The algorithm then performs a backward scan from the peak to locate the threshold crossing point, using linear interpolation between the two bracketing discrete. The overall pipeline is summarised in Figure~\ref{fig:pulse_formation}.

\section{Validation}\label{sec:validation}
A Monte Carlo-to-Monte Carlo (MC-to-MC) validation procedure has been designed to verify that the parametrised optical transport method produces output compatible with a full Geant4 ray-tracing simulation. The comparison is carried out at increasing levels of abstraction: first on the arrival-time distributions of the individual optical photons, then on the total number of photons extracted from the modules, and finally on the reconstructed figures of merit, i.e. the energy and time resolutions. Since the two simulation approaches share the identical digitisation and signal-formation chain described in Section~\ref{sec:digitisation}, any difference observed at any of these levels isolates the effect of the parametrised optical transport.

\subsection{Event-by-event comparison procedure}\label{sec:val_procedure}
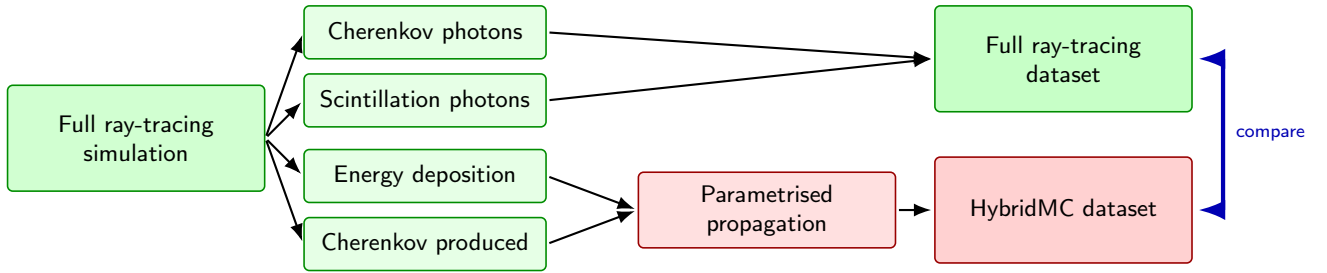
\begin{figure*}[pos=tbp]
\centering
\begin{tikzpicture}[
    font=\small, >=Latex,
    node distance=3mm and 5mm,
    box/.style={
        rectangle, rounded corners=2pt,
        draw, line width=0.6pt,
        minimum height=8mm, minimum width=34mm,
        align=center, inner sep=2pt
    },
    src/.style    ={box, fill=green!18, draw=green!55!black, minimum height=14mm},
    interm/.style    ={box, fill=green!10, draw=green!55!black, minimum width=32mm, minimum height=7mm},
    rtds/.style   ={box, fill=green!22, draw=green!55!black, minimum height=14mm},
    proc/.style   ={box, fill=red!12,   draw=red!60!black,   minimum height=10mm},
    hmcds/.style  ={box, fill=red!18,   draw=red!60!black,   minimum height=14mm},
    flow/.style   ={->, thick, shorten >=1pt, shorten <=1pt},
    cmp/.style    ={<->, ultra thick, blue!70!black, shorten >=2pt, shorten <=2pt}
]

\node[src] (src) at (0,0) {Full ray-tracing\\simulation};

\node[interm, right=of src, yshift=14mm]  (cer)  {Cherenkov photons};
\node[interm, right=of src, yshift=5mm]   (scn)  {Scintillation photons};
\node[interm, right=of src, yshift=-5mm]  (edep) {Energy deposition};
\node[interm, right=of src, yshift=-14mm] (cgen) {Cherenkov produced};

\node[proc,  right=12mm of edep, yshift=-4.5mm] (proc) {Parametrised\\propagation};
\node[rtds,  right=of proc,  yshift=20mm] (rtds) {Full ray-tracing\\dataset};
\node[hmcds, right=of proc]                     (hmds) {HybridMC dataset};

\draw[flow] (src.east) -- (cer.west);
\draw[flow] (src.east) -- (scn.west);
\draw[flow] (src.east) -- (edep.west);
\draw[flow] (src.east) -- (cgen.west);

\draw[flow] (cer.east) -- (rtds.west);
\draw[flow] (scn.east) -- (rtds.west);

\draw[flow] (edep.east) -- (proc.west);
\draw[flow] (cgen.east) -- (proc.west);
\draw[flow] (proc.east) -- (hmds.west);

\coordinate (cmpTR) at ([xshift=4mm]rtds.east);
\coordinate (cmpBR) at (cmpTR |- hmds.east);
\draw[cmp] (rtds.east) -- (cmpTR) -- (cmpBR)
   node[midway, right=1pt, font=\scriptsize]{compare}
   -- (hmds.east);

\end{tikzpicture}
\caption{Schematic representation of the Monte Carlo-to-Monte Carlo validation procedure. A single primary event is processed twice within the same full ray-tracing simulation: the scintillation and Cherenkov photons reaching the photo-detector volumes form the reference \emph{Full ray-tracing dataset}, while the energy-deposition map and the list of Cherenkov photons produced by the same event are also extracted and fed to the parametrised optical-transport step, yielding a second \emph{HybridMC dataset}. Comparing the two datasets event-by-event verifies that the parametrisation reproduces the full ray-tracing output.}\label{fig:validation_scheme}
\end{figure*}

The first two levels of the validation, on the photon arrival times and on the photon yield, rely on an event-by-event comparison whose procedure is schematically shown in Figure \ref{fig:validation_scheme}.
A full ray-tracing simulation mode is employed, wherein Geant4 is permitted to fully generate, propagate, and record a dataset of all scintillation and Cherenkov photons reaching the photodetector volumes. At the same time, the simulation still records the map of energy deposition and the list of Cherenkov photons generated, as it does in the standard HybridMC workflow. This information is used to perform the operations described in Section~\ref{sec:pot} to obtain a second dataset of photons reaching the photodetector volumes. These two datasets are then compared to verify their compatibility. Because the same primary event is processed in both modes, this approach provides an event-by-event test of the parametrisation. This dataset is used to perform two photon-level comparisons: one on the arrival-time distributions (Section~\ref{sec:val_photon}) and one on the photon yield (Section~\ref{sec:val_yield}).
The validation focuses on two representative module configurations, one per technology supported by the framework: a longitudinally segmented SpaCal module with tungsten absorber and GAGG crystals fibres (W-GAGG), and a longitudinally segmented Shashlik module. For the event-by-event comparison, a total of 1000 primary electrons of 1~GeV/c momentum were shot at the centre of each module.

\subsection{Photon arrival-time distributions}\label{sec:val_photon}
\begin{figure*}[pos=tbp]
  \centering
  \begin{subfigure}[t]{0.49\textwidth}
    \includegraphics[width=\linewidth]{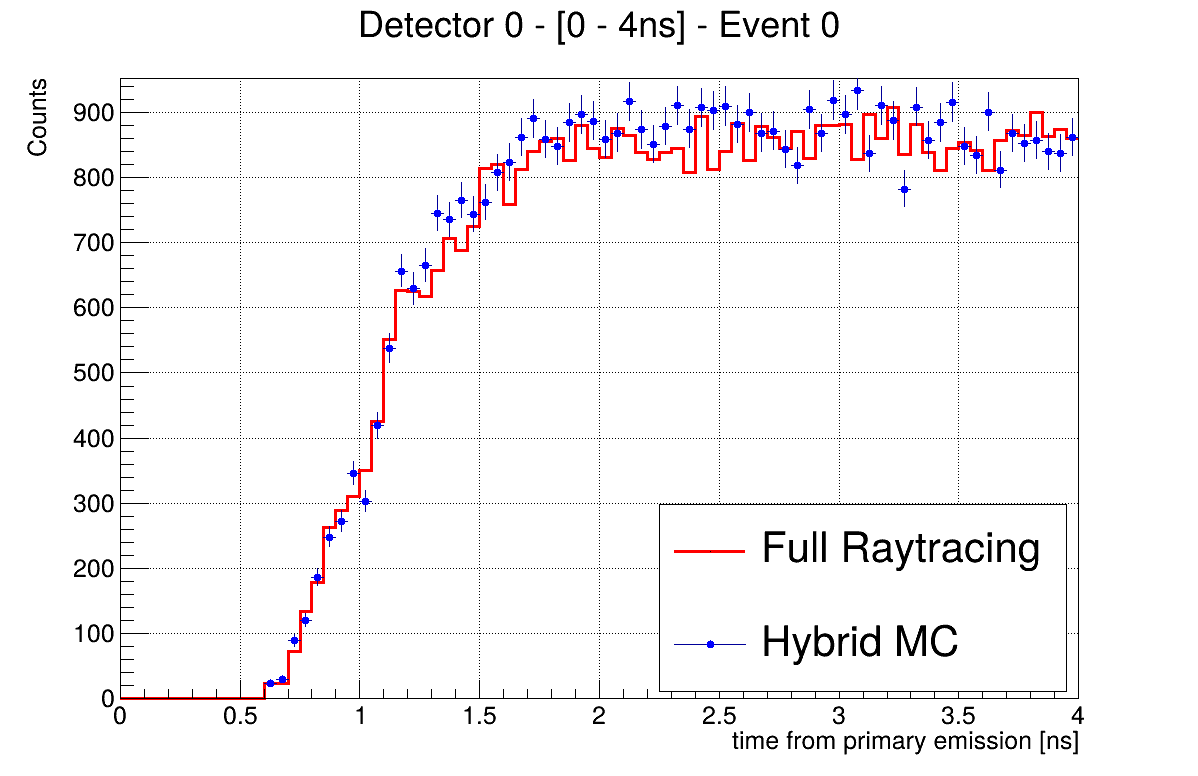}
    \caption{Front readout, leading edge.}
  \end{subfigure}\hfill
  \begin{subfigure}[t]{0.49\textwidth}
    \includegraphics[width=\linewidth]{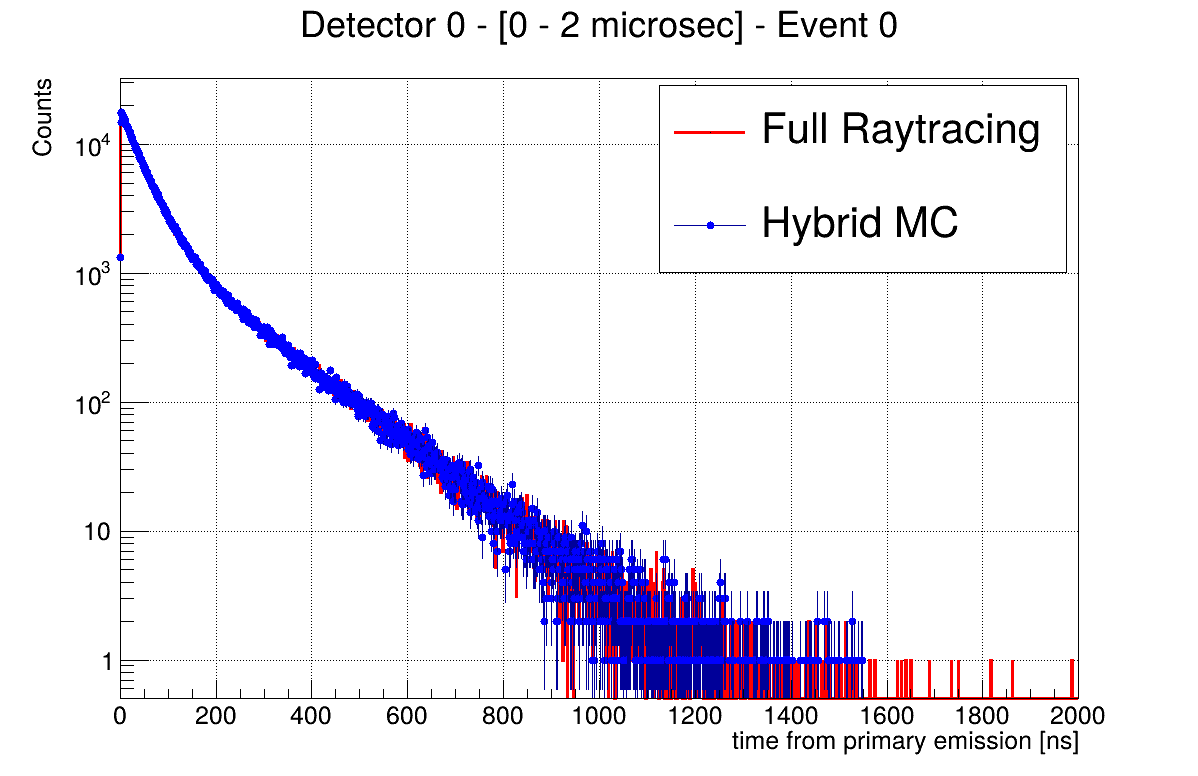}
    \caption{Front readout, full pulse.}
  \end{subfigure}
  \\[4pt]
  \begin{subfigure}[t]{0.49\textwidth}
    \includegraphics[width=\linewidth]{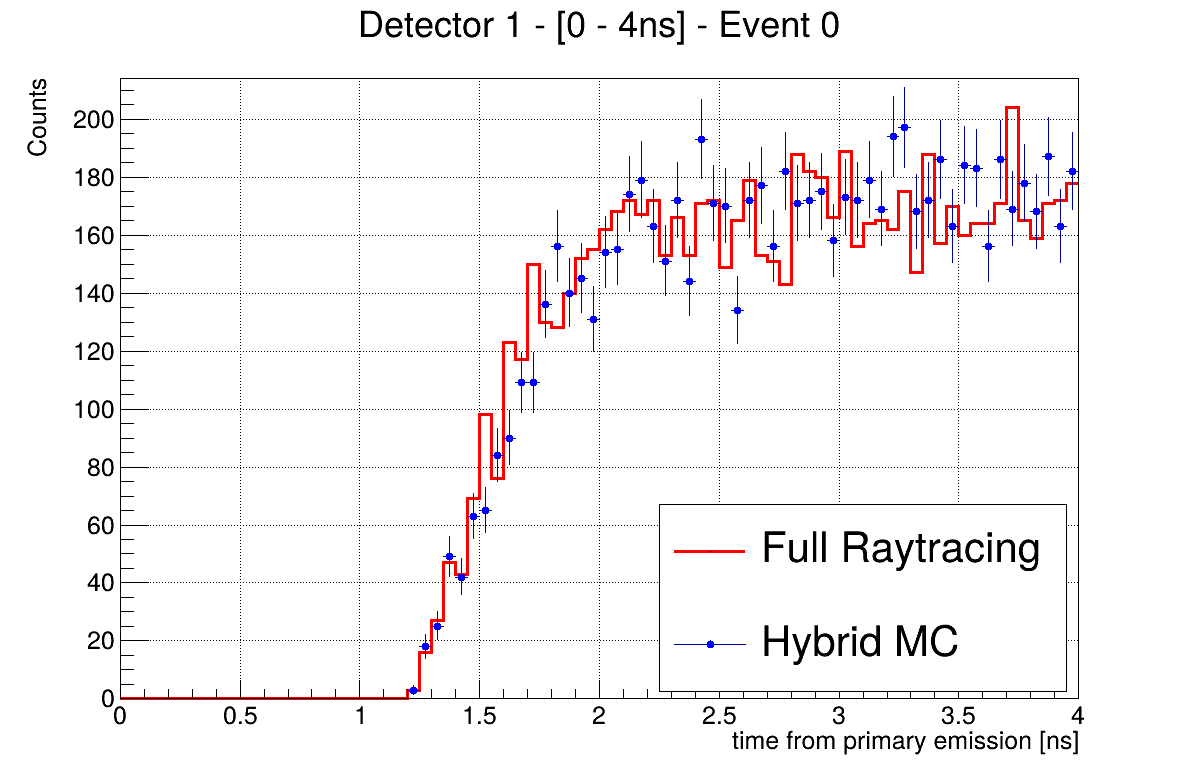}
    \caption{Back readout, leading edge.}
  \end{subfigure}\hfill
  \begin{subfigure}[t]{0.49\textwidth}
    \includegraphics[width=\linewidth]{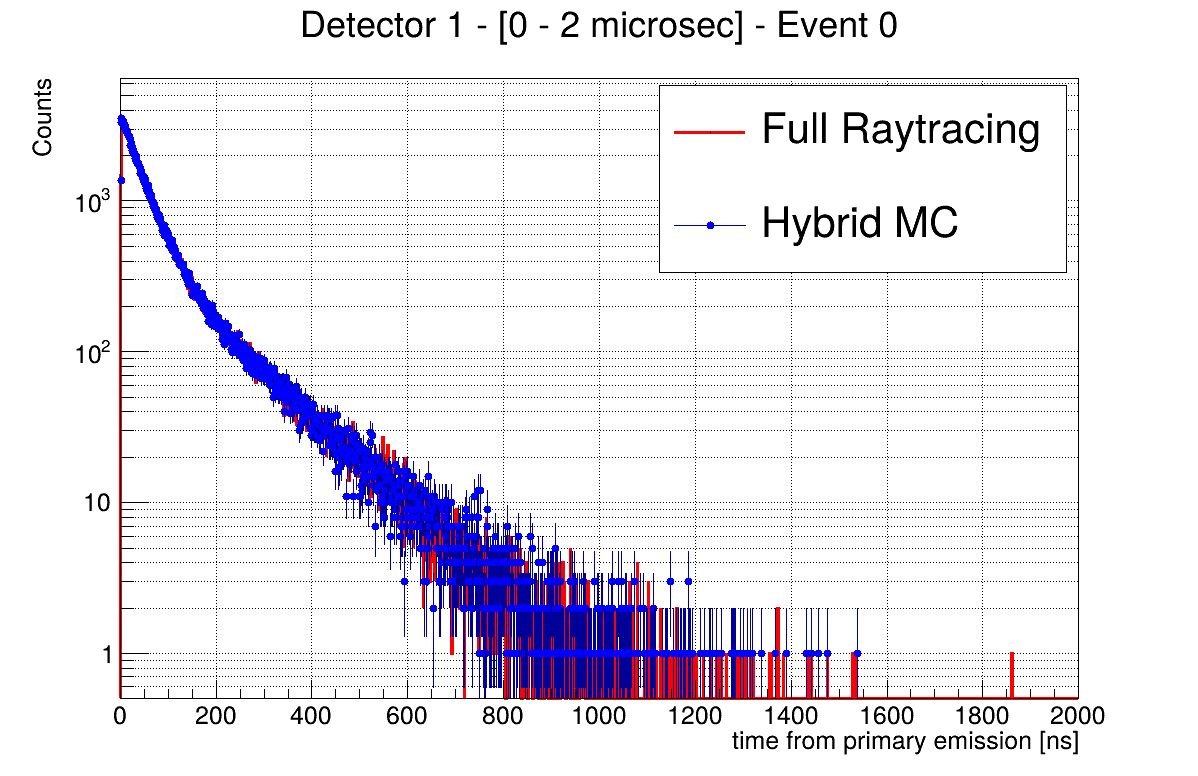}
    \caption{Back readout, full pulse.}
  \end{subfigure}
  \caption{Time-of-arrival distributions of optical photons at the photo-detector volumes of a SpaCal W-GAGG module for a $1$~GeV$/c$ primary electron, comparing full Geant4 ray-tracing (solid) with HybridMC parametrised optical transport (dashed). A single example event is shown out of the 1000 simulated for this module type. The statistical comparison over the full sample is reported in Figure~\ref{fig:ks_distribution}. Top row: front readout face (electron-entrance side). Bottom row: back readout face. Left column: zoom on the leading edge. Right column: full time window.}\label{fig:validation_spacal_w_gagg}
\end{figure*}

\begin{figure*}[pos=tbp]
  \centering
  \begin{subfigure}[t]{0.49\textwidth}
    \includegraphics[width=\linewidth]{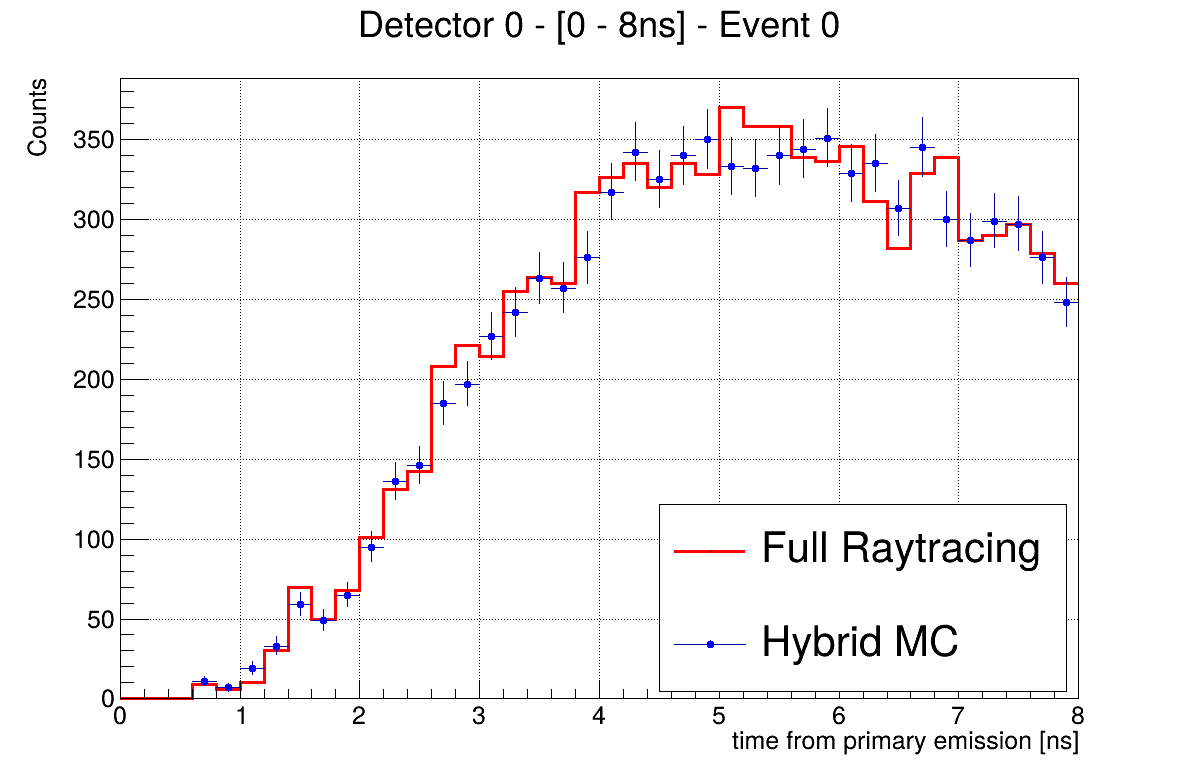}
    \caption{Front readout, leading edge.}
  \end{subfigure}\hfill
  \begin{subfigure}[t]{0.49\textwidth}
    \includegraphics[width=\linewidth]{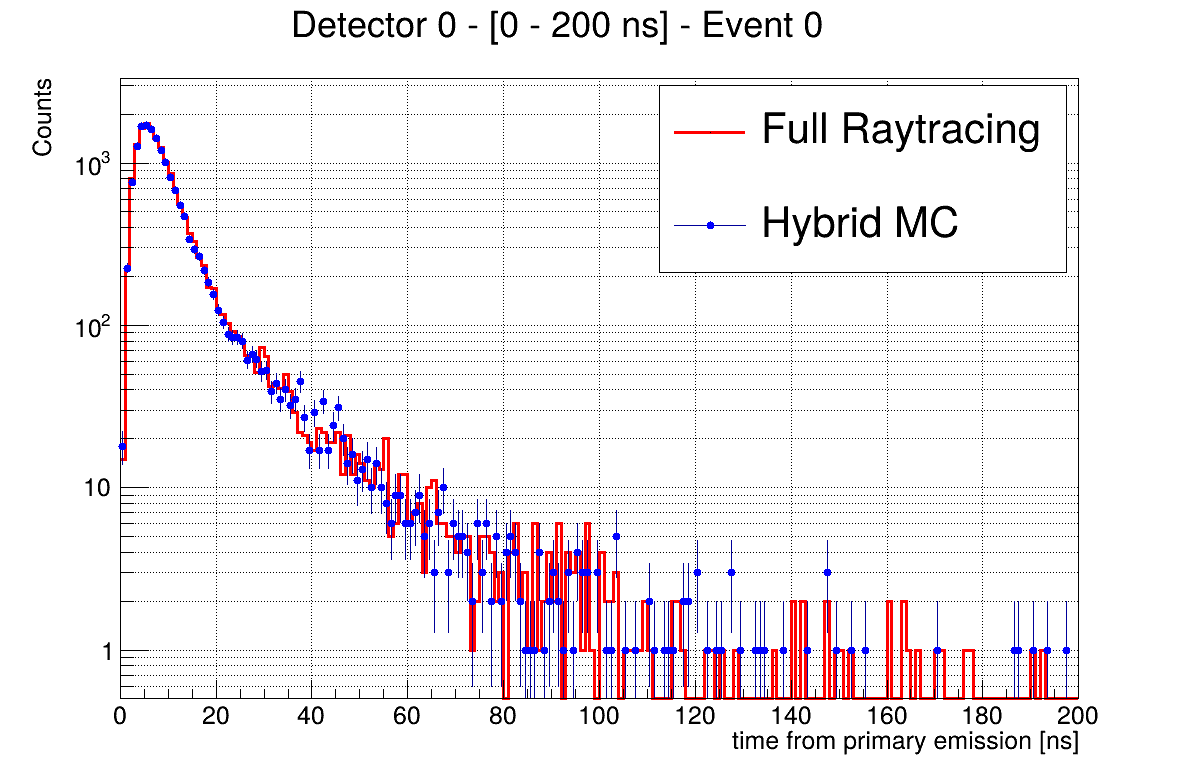}
    \caption{Front readout, full pulse.}
  \end{subfigure}
  \\[4pt]
  \begin{subfigure}[t]{0.49\textwidth}
    \includegraphics[width=\linewidth]{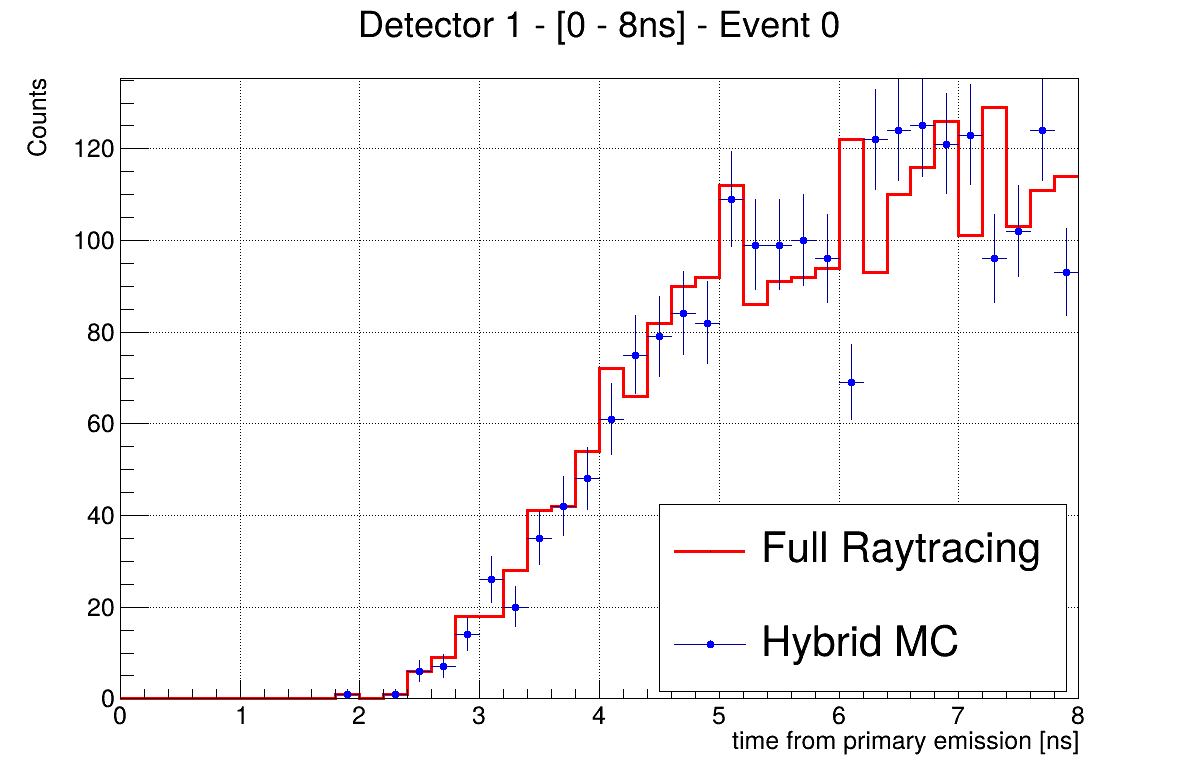}
    \caption{Back readout, leading edge.}
  \end{subfigure}\hfill
  \begin{subfigure}[t]{0.49\textwidth}
    \includegraphics[width=\linewidth]{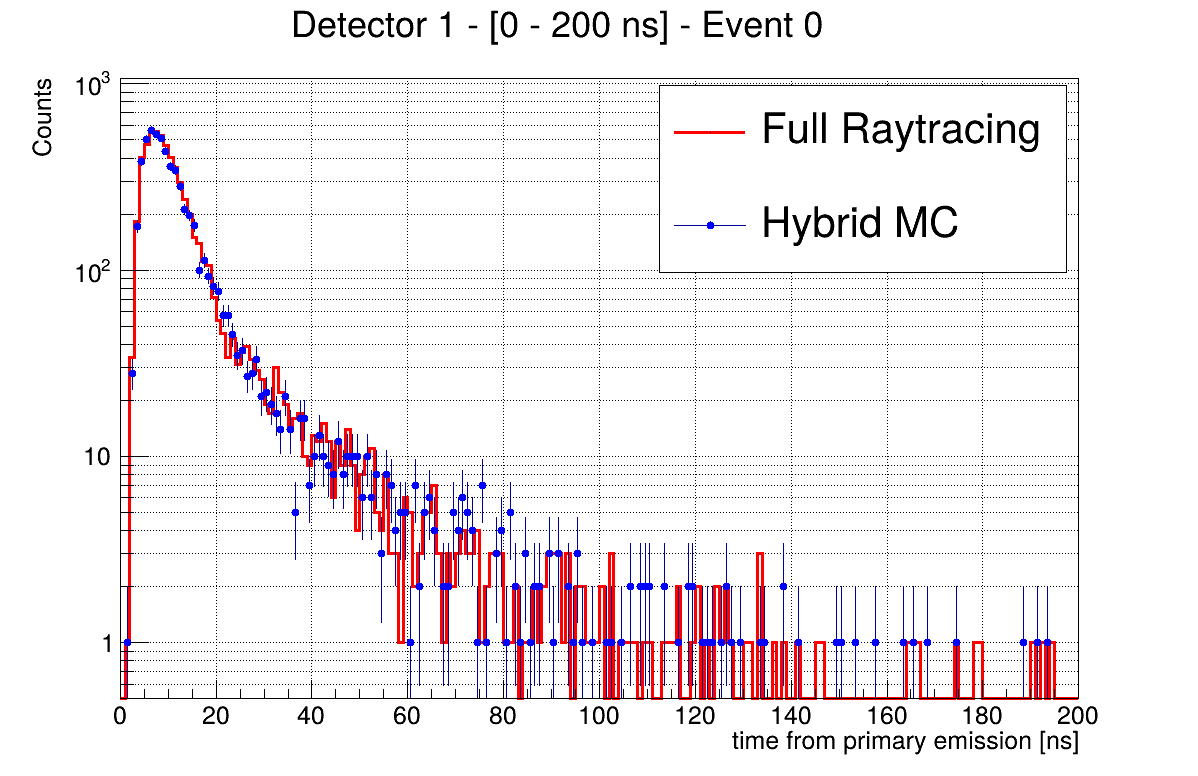}
    \caption{Back readout, full pulse.}
  \end{subfigure}
  \caption{Time-of-arrival distributions of optical photons at the photo-detector volumes of a Shashlik module for a $1$~GeV$/c$ primary electron, comparing full Geant4 ray-tracing (solid) with HybridMC parametrised optical transport (dashed). A single example event is shown out of the 1000 simulated for this module type. The statistical comparison over the full sample is reported in Figure~\ref{fig:ks_distribution}. Top row: front readout face (electron-entrance side). Bottom row: back readout face. Left column: zoom on the leading edge. Right column: full time window.}\label{fig:validation_shashlik}
\end{figure*}

The distributions of the time of arrival of optical photons obtained are reported for the SpaCal W-GAGG (Figure \ref{fig:validation_spacal_w_gagg}) and Shashlik (Figure \ref{fig:validation_shashlik}) modules. In both figures the distributions refer to a single example event out of the 1000 simulated for each module type. The agreement between the full ray-tracing and the parametrised transport is visually evident, both in the leading edge of the distribution (left panels) and in the long tail (right panels).

\begin{figure*}[pos=tbp]
  \centering
  \begin{subfigure}[t]{0.49\textwidth}
    \includegraphics[width=\linewidth]{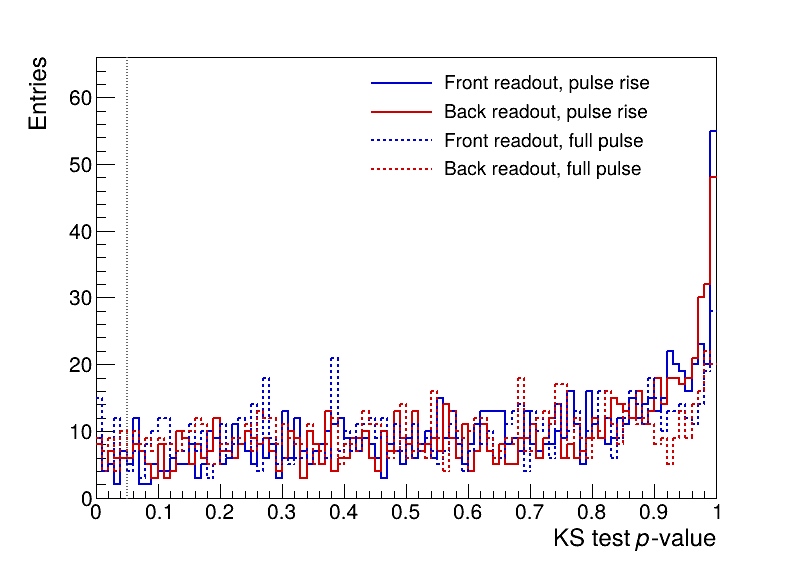}
    \caption{SpaCal W-GAGG.}
  \end{subfigure}\hfill
  \begin{subfigure}[t]{0.49\textwidth}
    \includegraphics[width=\linewidth]{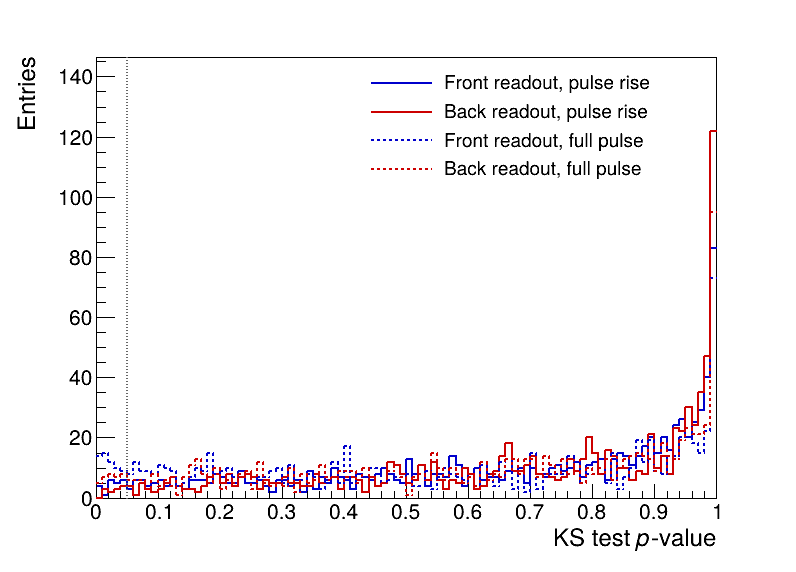}
    \caption{Shashlik.}
  \end{subfigure}
  \caption{Distribution of the Kolmogorov\textendash Smirnov $p$-values obtained by comparing, on a pulse-by-pulse basis, the full ray-tracing reference dataset to the HybridMC parametrised optical-transport dataset, for the SpaCal W-GAGG (left) and Shashlik (right) module technologies. Each panel overlays the four readout-window combinations: front (blue) and back (red) readout faces, for the leading-edge (solid) and full-pulse (dashed) time windows. The dotted vertical line marks the conventional $p=0.05$ acceptance threshold. The distributions are approximately flat with a pronounced pileup towards $p=1$ and only a small fraction of entries below the threshold, as quantified in Table~\ref{table:ks_summary}.}
  \label{fig:ks_distribution}
\end{figure*}

To quantitatively assess the statistical equivalence of these temporal distributions, the Kolmogorov-Smirnov (KS) test is applied across all of the individual simulated photodetector pulses. The distributions of the resulting $p$-values, reported in Figure~\ref{fig:ks_distribution} for the W-GAGG and Shashlik module technologies, are approximately uniform with a marked pileup towards $p=1$ and only a small fraction of entries below the conventional $p=0.05$ threshold: good agreement, above the threshold, was found for the large majority of the dataset across both module technologies and all readout configurations (see Table~\ref{table:ks_summary}).

\begin{table}[pos=tbp]
\caption{Summary of the Kolmogorov\textendash Smirnov agreement test between the full ray-tracing reference dataset and the HybridMC parametrised optical-transport dataset, broken down by module technology and readout channel. The reported values are the fractions of simulated photo-detector pulses for which the KS test yields $p>0.05$, for the leading-edge and full-pulse time windows and for the two readout detectors (Front and Back).}\label{table:ks_summary}
\begin{tabular*}{\tblwidth}{@{}LCCCC@{}}
\toprule
 & \multicolumn{2}{c}{\textbf{Leading edge}} & \multicolumn{2}{c}{\textbf{Full pulse}} \\
\cmidrule(lr){2-3}\cmidrule(lr){4-5}
\textbf{Module technology} & \textbf{Front} & \textbf{Back} & \textbf{Front} & \textbf{Back} \\
\midrule
SpaCal W/GAGG & 97.2\% & 96.8\% & 95.6\% & 96.0\% \\
Shashlik      & 97.7\% & 98.8\% & 93.8\% & 96.4\% \\
\bottomrule
\end{tabular*}
\end{table}

\subsection{Photon yield}\label{sec:val_yield}
\begin{figure*}[pos=tbp]
  \centering
  \begin{subfigure}[t]{0.49\textwidth}
    \includegraphics[width=\linewidth]{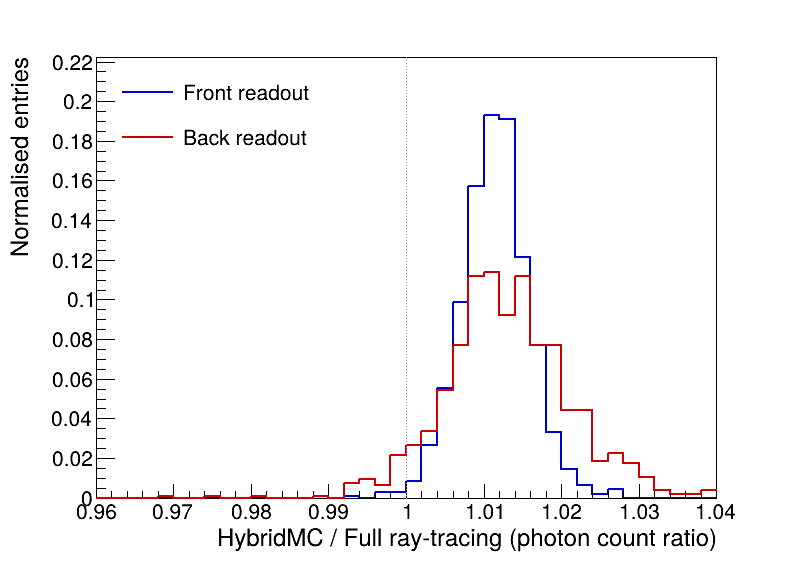}
    \caption{SpaCal W-GAGG.}\label{fig:photon_ratio_wgagg}
  \end{subfigure}\hfill
  \begin{subfigure}[t]{0.49\textwidth}
    \includegraphics[width=\linewidth]{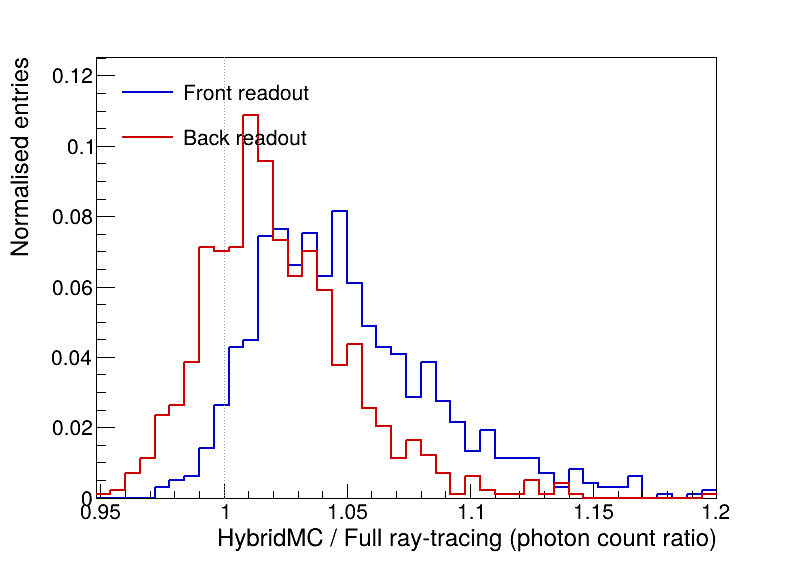}
    \caption{Shashlik.}\label{fig:photon_ratio_shashlik}
  \end{subfigure}
  \caption{Distribution of the ratio between the number of optical photons extracted with the HybridMC parametrised optical transport and with the full ray-tracing reference, evaluated event-by-event for the front (electron-entrance side, blue) and back (red) readout faces, for the SpaCal W-GAGG (left) and Shashlik (right) module technologies. The dotted line marks the unity ratio. Note the different horizontal scales of the two panels.}\label{fig:photon_ratio}
\end{figure*}

Beyond the temporal shape of the pulses, it is important to assess the consistency of the hybrid approach in terms of the number of optical photons extracted from the modules, as the amount of collected light has a direct impact on both the energy and the time resolution of the calorimeter. %
The total number of photons reaching each readout face was computed on an event-by-event basis, and the ratio between the HybridMC and the full ray-tracing values was histogrammed. Figure~\ref{fig:photon_ratio} shows these distributions for the SpaCal W-GAGG and the Shashlik modules, separately for the front (electron-entrance side) and back readout faces. In both technologies the ratio distributions are centred close to unity, but a slight positive bias of a few percent is observed.
The W-GAGG distributions are sharply peaked within a few percent of unity, while the Shashlik module exhibits a broader spread. The origin of this difference is discussed in Section~\ref{sec:discussion}.

\subsection{Energy and time resolution}\label{sec:val_observables}
The agreement demonstrated at the level of individual photons must ultimately translate into the figures of merit that drive the detector design: the energy resolution and the time resolution of the calorimeter modules. This final level of the validation differs from the event-by-event comparison of Sections~\ref{sec:val_photon} and~\ref{sec:val_yield}: rather than processing the same primary event in both modes, two independent simulation campaigns were carried out on each of the two representative modules, one entirely in full ray-tracing mode and one entirely in HybridMC mode, with primary electrons spanning the $1$--$100$~GeV/c momentum range. This range covers the typical operating conditions of the calorimeter and, spanning two decades in momentum, gives access to both the sampling and the constant term of the energy resolution. The energy- and time-resolution curves obtained from the two campaigns were then compared. Both campaigns are processed through the identical digitisation and signal-formation chain of Section~\ref{sec:digitisation}, so that any difference between the reconstructed figures of merit isolates the effect of the optical-transport parametrisation.

\begin{figure*}[pos=tbp]
  \centering
  \begin{subfigure}[b]{0.49\textwidth}
    \includegraphics[width=\linewidth]{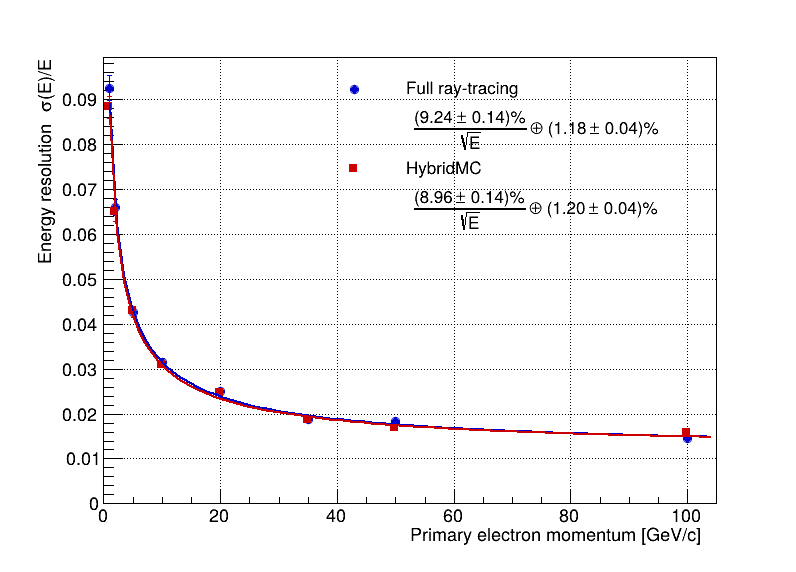}
    \caption{SpaCal W-GAGG.}
  \end{subfigure}\hfill
  \begin{subfigure}[b]{0.49\textwidth}
    \includegraphics[width=\linewidth]{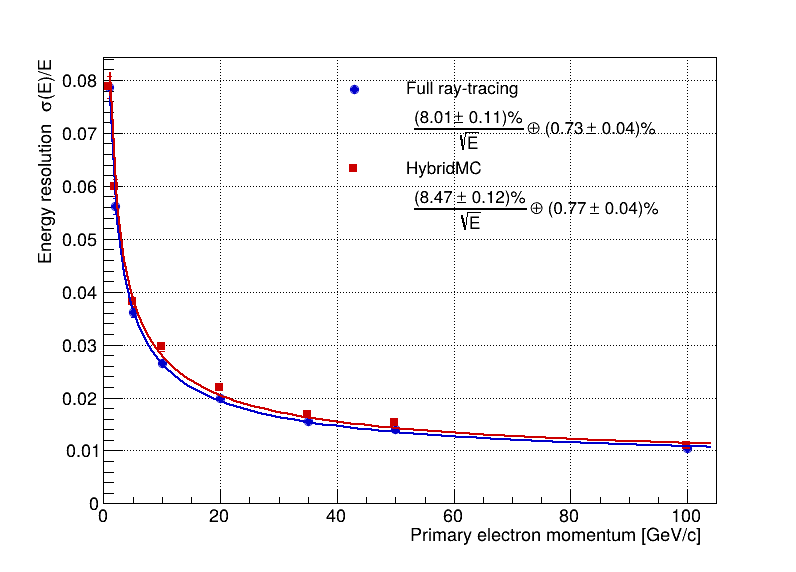}
    \caption{Shashlik.}
  \end{subfigure}
  \caption{Energy resolution as a function of the primary-electron momentum for the SpaCal W-GAGG (left) and Shashlik (right) modules, obtained with full Geant4 ray-tracing (blue) and with the HybridMC parametrised optical transport (red). Both datasets are processed through the identical digitisation and signal-formation chain. The curves are fits with $\sigma(E)/E = a/\sqrt{E} \oplus b$.}\label{fig:val_eres}
\end{figure*}

The energy resolution is compared in Figure~\ref{fig:val_eres}. For the W-GAGG module, fitting both curves with the standard parametrisation $\sigma(E)/E = a/\sqrt{E} \oplus b$ yields a sampling term $a = (9.24 \pm 0.14)\%$ and a constant term $b = (1.18 \pm 0.04)\%$ for the full ray-tracing reference, against $a = (8.96 \pm 0.14)\%$ and $b = (1.20 \pm 0.04)\%$ for HybridMC: the two sets of fitted terms are compatible within the statistical uncertainties over the entire two-decade momentum range. For the Shashlik module the fit gives $a = (8.01 \pm 0.11)\%$ and $b = (0.73 \pm 0.04)\%$ for the full ray-tracing reference, against $a = (8.47 \pm 0.12)\%$ and $b = (0.77 \pm 0.04)\%$ for HybridMC. The two sets of fitted terms are reasonably compatible within their statistical uncertainties, although the HybridMC curve lies slightly above the reference at every momentum point. The origin of this systematic offset is discussed in Section~\ref{sec:discussion}.

\begin{figure*}[pos=tbp]
  \centering
  \begin{subfigure}[b]{0.49\textwidth}
    \includegraphics[width=\linewidth]{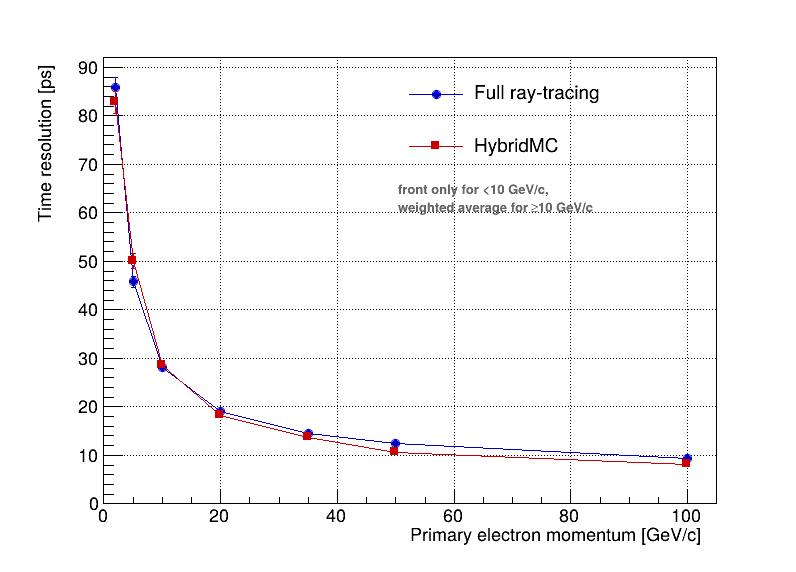}
    \caption{SpaCal W-GAGG.}
  \end{subfigure}\hfill
  \begin{subfigure}[b]{0.49\textwidth}
    \includegraphics[width=\linewidth]{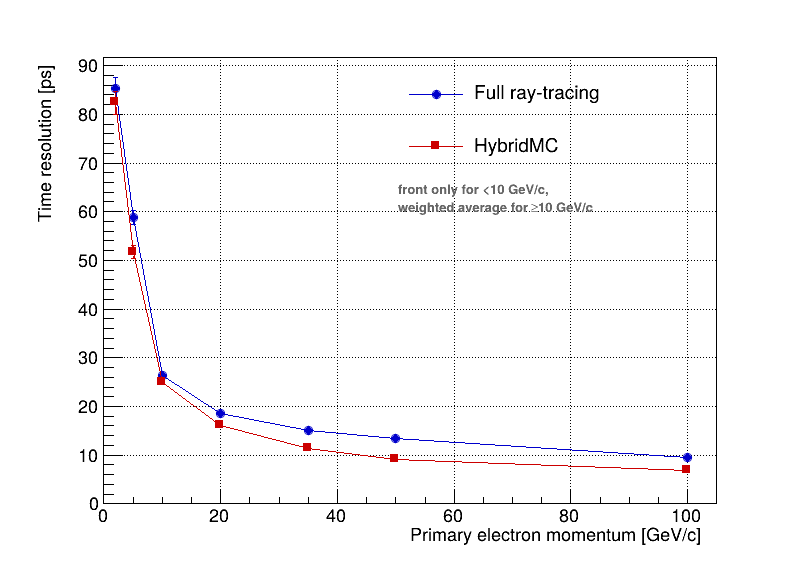}
    \caption{Shashlik.}
  \end{subfigure}
  \caption{Time resolution as a function of the primary-electron momentum for the SpaCal W-GAGG (left) and Shashlik (right) modules, obtained with full Geant4 ray-tracing (blue) and with the HybridMC parametrised optical transport (red). The resolution is taken from the front readout section below $10$~GeV/c and from the front/back weighted average above it. Both datasets are processed through the identical digitisation and signal-formation chain.}\label{fig:val_tres}
\end{figure*}

The time resolution is compared in Figure~\ref{fig:val_tres}, where the curve is built from the front-section signal at low momentum and from the front/back weighted average above $10$~GeV/c. Only the front section is used at low momentum because, in the configurations tested, the signal collected in the back section is too small there to allow a reliable measurement of the time resolution. Above $10$~GeV/c the back-section signal becomes sufficient and the two readouts are combined into a weighted average. For the W-GAGG module the two simulations track each other from about $85$~ps at $2$~GeV/c down to the $10$~ps level at the highest momenta. For the Shashlik module the two curves follow the same trend, from about $85$~ps at $2$~GeV/c down to below $10$~ps at the highest momenta, but HybridMC lies systematically below the full ray-tracing reference, by a few percent at $2$~GeV/c and by close to thirty percent at $100$~GeV/c. The parametrised transport therefore reproduces the time resolution of the reference for the W-GAGG module over the whole momentum range, while for the Shashlik module it tends to predict a somewhat better resolution than the reference. The origin of this behaviour, and its connection with the photon yield of Section~\ref{sec:val_yield}, is discussed in Section~\ref{sec:discussion}.

\section{Gain in computation time}\label{sec:gain}

The performance gain provided by the parametrised optical-transport method has been evaluated by comparing HybridMC simulations with simulations performed with the same framework but employing full ray-tracing. Acceleration factors of more than two orders of magnitude, heavily dependent on the specific scintillator material and module geometry, have been found, as reported in Table \ref{table:speedup}.
\begin{table*}[pos=tbp]
\caption{CPU time for different module technologies, measured in the same condition for full ray-tracing simulations and using the HybridMC method. The CPU times are expressed in seconds per GeV of energy of the primary particle. All quoted values are inverse-variance weighted averages over the available momenta, with PDG-style scaled uncertainties. The W-GAGG and Shashlik configurations were scanned over the full $1$--$100$~GeV/c range, whereas the Pb-Polystyrene and W-Polystyrene comparison is restricted to the $1$--$10$~GeV/c range ($1$, $3$, $5$ and $10$~GeV/c), which is justified by the first-order flatness of the speed-up shown in Figure~\ref{fig:cputime_flatness}.}\label{table:speedup}
\begin{tabular*}{\tblwidth}{@{}CCCCC@{}}
\toprule
 \textbf{Technology} & \textbf{Type} & \textbf{Full Ray-tracing} & \textbf{HybridMC} & \textbf{Gain Factor} \\ %
            &      & [s/GeV] & [s/GeV] & \\ %

\midrule
\multirow{3}{*}{SpaCal} & W-GAGG & 1146 $\pm$ 84 & 7.30 $\pm$ 0.27 & 157 $\pm$ 13 \\
&Pb-Polystyrene & 2113 $\pm$ 179 & 4.40 $\pm$ 0.63 & 481 $\pm$ 80 \\
&W-Polystyrene & 879 $\pm$ 26 & 3.01 $\pm$ 0.24 & 292 $\pm$ 25 \\
Shashlik & - & 1033 $\pm$ 24 & 3.67 $\pm$ 0.08 & 281 $\pm$ 9 \\
\bottomrule

\end{tabular*}
\end{table*}

The gain in computation time for the W-GAGG SpaCal and the Shashlik modules has been evaluated over the full $1$--$100$~GeV/c momentum range, using the same dataset produced for the study described in Section~\ref{sec:val_observables}. The CPU time of a job is taken as the sum of the user and system time charged to it by the batch system, so that the input/output overhead of loading the optical calibration tables is accounted for. At each momentum, the CPU time of each approach was normalised to the number of primary particles simulated per job and to the primary-particle momentum, yielding the CPU time per primary particle and per GeV. The inverse-variance weighted average of these measurements over all momenta, with a PDG-style scale factor accounting for their spread~\cite{PDG2024}, gives the corresponding entries of Table~\ref{table:speedup}, namely an overall speed-up of $157\pm13$ for the W-GAGG module and $281\pm9$ for the Shashlik module with respect to full ray-tracing.

\begin{figure}[pos=tbp]
  \centering
    \includegraphics[width=\linewidth]{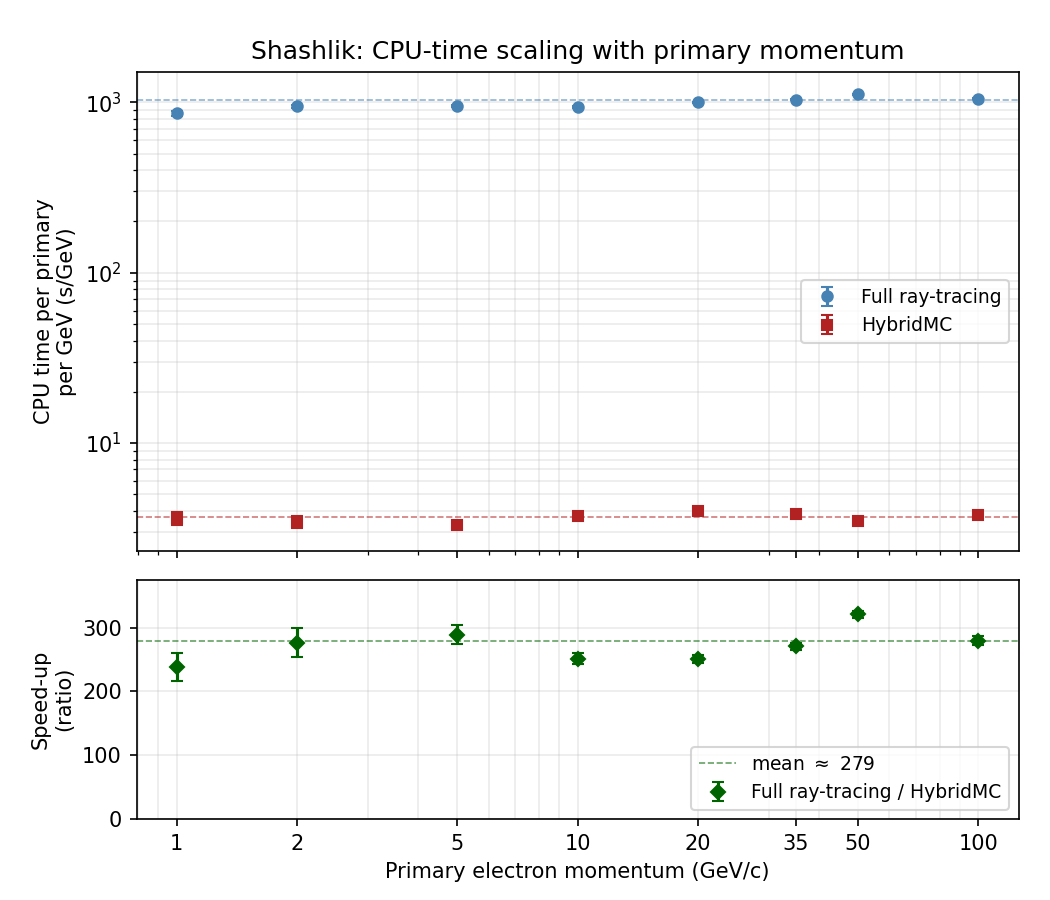}
    \caption{CPU-time scaling with the primary-electron momentum for the Shashlik module. \textbf{Top:} CPU time per primary particle and per GeV for the full ray-tracing and the HybridMC approaches, on a logarithmic scale (the two differ by about two and a half decades). The dashed lines mark the respective weighted means. \textbf{Bottom:} their ratio, i.e. the HybridMC speed-up. Both the per-GeV CPU cost and the ratio remain flat, to first order, over the $1$--$100$~GeV/c range.}\label{fig:cputime_flatness}
\end{figure}

For both modules, the CPU time per primary particle and per GeV is found to remain, to first order, flat over the whole momentum range, for the full ray-tracing as well as for the HybridMC approach, and so is their ratio. This is illustrated for the Shashlik module in Figure~\ref{fig:cputime_flatness}: across the two and a half decades spanned by the two simulation approaches, neither the per-GeV CPU cost nor the speed-up shows any significant trend with the primary-particle momentum. The same behaviour is observed for the W-GAGG module. This is expected, since the CPU cost of both approaches scales, to first order, with the deposited energy.

This first-order flatness justifies characterising the speed-up over a restricted momentum range when a full scan is too expensive. For the Pb-Polystyrene and W-Polystyrene configurations, for which a full ray-tracing scan is particularly demanding in CPU time (especially for Pb-Polystyrene), the comparison was therefore performed over the restricted $1$--$10$~GeV/c range, sampled at $1$, $3$, $5$ and $10$~GeV/c. Given the flatness established above, the speed-up measured over this restricted range is taken as representative of the full momentum range.

\begin{figure}[pos=tbp]
  \centering
    \includegraphics[width=\linewidth]{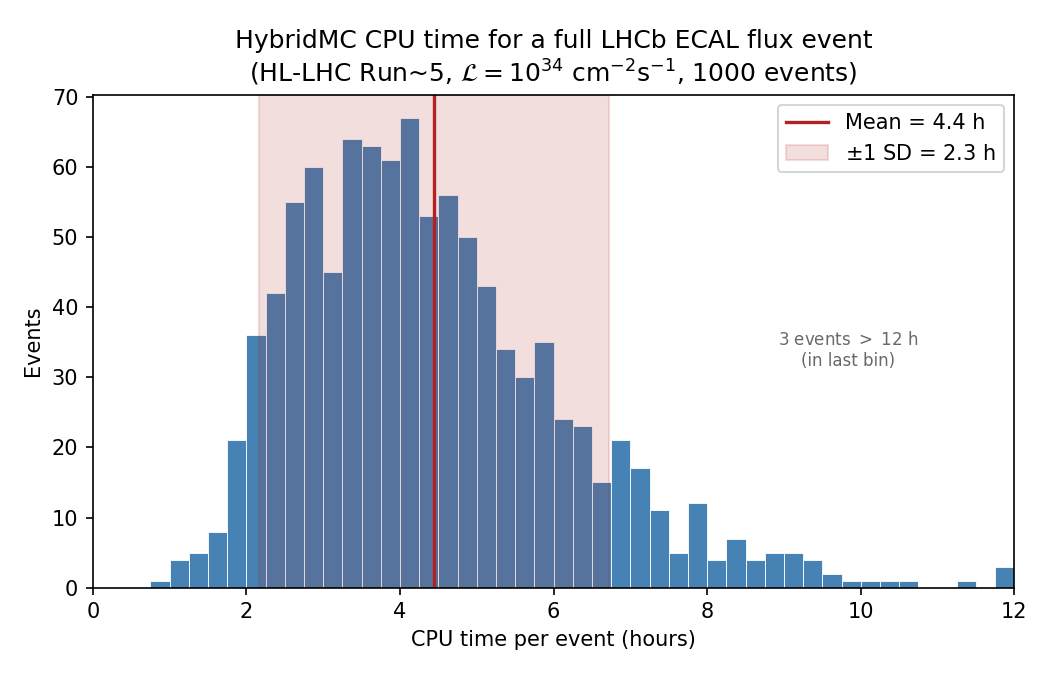}
    \caption{Distribution of the CPU time per event required by HybridMC to simulate the full LHCb ECAL response to the minimum-bias particle flux expected at the HL-LHC Run~5 luminosity of $\mathcal{L}=1.0\times10^{34}$~cm$^{-2}$s$^{-1}$, for a sample of $1000$ events. The solid line marks the mean ($4.4$~h) and the shaded band the $\pm1$ standard deviation ($2.3$~h). The two events exceeding $12$~h are collected in the last bin.}\label{fig:flux_cpu_time}
\end{figure}

Full-scale LHCb ECAL simulations incorporate a mixture of SpaCal and Shashlik technologies. In this configuration, the full flux of particles arriving at the ECAL after a proton-proton collision needs to be simulated. To assess the cost of such simulations, a sample of $1000$ minimum-bias events, generated with the LHCb Simulation Software~\cite{CERN-LHCC-2018-007}, was processed with HybridMC through the complete upgraded ECAL geometry, under the particle flux expected at the HL-LHC Run~5 instantaneous luminosity of $\mathcal{L}=1.0\times10^{34}$~cm$^{-2}$s$^{-1}$. The resulting distribution of the CPU time per event is shown in Figure~\ref{fig:flux_cpu_time}: it is peaked around $4$~hours with a tail towards larger values, giving a mean of $(4.4\pm2.3)$~hours per bunch crossing.
The implications of this figure for the feasibility of full-scale simulations are discussed in Section~\ref{sec:discussion}.

\section{Comparison to test-beam data}\label{sec:comparison}

A comparison to experimental data is reported here to assess whether the description of a real module implemented in the framework is adequate. This is a question distinct from the validation of the method, which is established entirely by the Monte Carlo-to-Monte Carlo comparison of Section~\ref{sec:validation}. Residual differences between data and simulation at this level can originate from characteristics of the experimental setup or of the physical prototype that are not reproduced by the idealised simulation, and do not reflect on the validity of the hybrid method.

The comparison is carried out on a recent W-GAGG SpaCal module, similar to the one used for the validation of Section~\ref{sec:validation}, built as a tungsten absorber embedded with scintillating crystal fibres. The geometry corresponds to the parameters listed in Table~\ref{table:configurations}, and the scintillator is an ultrafast, radiation-hard GAGG developed for the R\&D of the LHCb ECAL Upgrade~II~\cite{Roux_GAGG_PicoCal}. A prototype of this module was characterised in 2025, with electrons at the DESY~II synchrotron in the $1$--$5$~GeV/c momentum range and at the CERN Super Proton Synchrotron (SPS) North Area in the $20$--$100$~GeV/c range, using picosecond time-reference detectors whose contribution is subtracted from the measured time resolution~\cite{An_2022_TB}. The same module was described in the framework using the configuration system of Section~\ref{sec:mcsim} and run through the complete HybridMC chain. Both the energy and the time resolution are compared.

The relative energy resolution $\sigma(E)/E$ and the time resolution $\sigma_t$ are each parametrised as $a/\sqrt{p}\oplus c$, where $p$ is the primary-electron momentum, $a$ the stochastic term, $c$ the constant term, and $\oplus$ denotes addition in quadrature. The simulated points are fitted with this function. To quantify the residual difference from the measurement, an additional constant term $\Delta c$ is then added in quadrature to the fitted simulation, giving $\sqrt{a^2/p + c^2 + \Delta c^2}$, and fitted as the only free parameter against the test-beam points, with $a$ and $c$ held at their simulated values. The corrected simulation is drawn as a band at $\pm2\sigma$, propagated from the covariance of the fit. Because $\Delta c$ is fitted against the data, it is not an independent prediction of the simulation. It measures, with a single momentum-independent number, what the simulation is missing.

\begin{figure*}[pos=tbp]
  \centering
  \begin{subfigure}[b]{0.49\textwidth}
    \includegraphics[width=\linewidth]{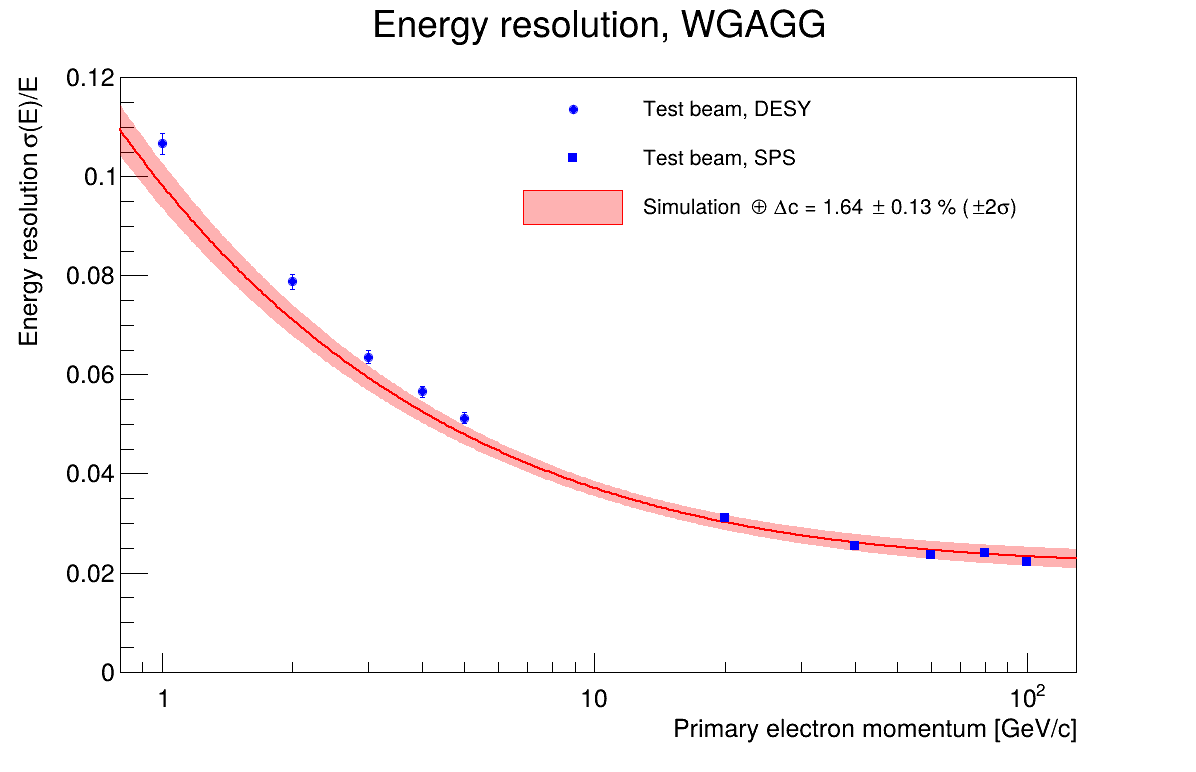}
    \caption{Energy resolution.}
  \end{subfigure}\hfill
  \begin{subfigure}[b]{0.49\textwidth}
    \includegraphics[width=\linewidth]{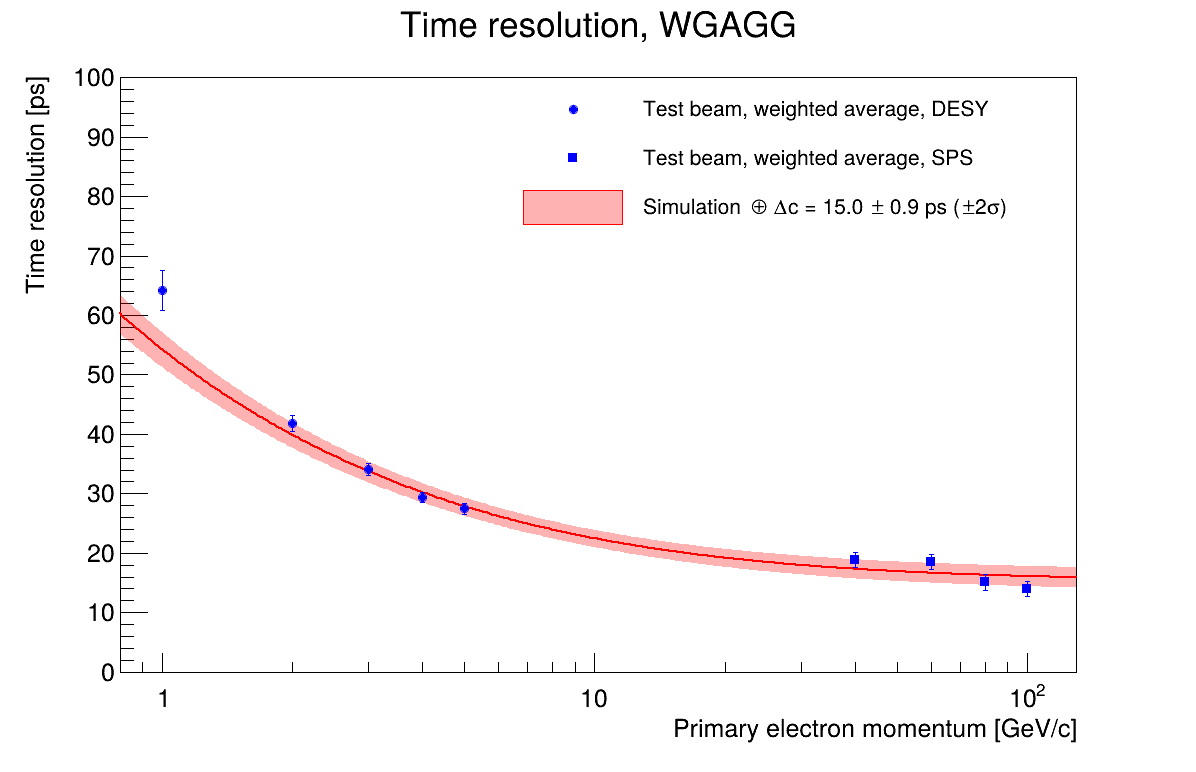}
    \caption{Time resolution.}
  \end{subfigure}
  \caption{Energy resolution (left) and time resolution (right) of the W-GAGG SpaCal module as a function of the primary-electron momentum, comparing the HybridMC simulation to test-beam measurements collected at DESY~II ($1$--$5$~GeV/c, circles) and at the CERN SPS ($20$--$100$~GeV/c, squares). The simulation is fitted with $a/\sqrt{p}\oplus c$, applied to $\sigma(E)/E$ for the energy resolution and to $\sigma_t$ for the time resolution, and the red band represents the results of the simulation plus an additional constant term $\Delta c$ fitted against the test-beam points and added in quadrature, drawn at $\pm2\sigma$. The fitted $\Delta c$ is $(1.64\pm0.13)\%$ for the energy resolution and $(15.0\pm0.9)$~ps for the time resolution. For the timing, the front-section signal is used below $4$~GeV/c and the front/back weighted average above it, and the SPS point at $20$~GeV/c is excluded as an outlier. The momentum axis is logarithmic.}\label{fig:tb_comparison}
\end{figure*}

The energy resolution is shown in Figure~\ref{fig:tb_comparison} (left). The simulated stochastic term, $a=(9.6\pm0.2)\%$, is close to the test-beam value of $(10.6\pm0.1)\%$, so the simulation reproduces the momentum dependence of the resolution over the full two-decade range. The residual difference is captured by a constant term $\Delta c = (1.64\pm0.13)\%$, added to the simulated constant term of $1.35\%$. A residual constant term of this size is expected for a real module: it collects the effects of mechanical tolerances, cell-to-cell calibration spread and inter-cell gaps that the idealised single-module simulation does not include.

The time resolution is shown in Figure~\ref{fig:tb_comparison} (right). %
The simulated stochastic term is $a=(52\pm2)$~ps\,$\sqrt{\mathrm{GeV}/c}$ with a constant term of $2.8$~ps, and the residual is a constant term $\Delta c = (15.0\pm0.9)$~ps. The resolution of the time-reference detectors is subtracted from the measured points, so this term instead reflects contributions present in the measurement but absent from the simulation, such as the trigger and clock jitter, the time calibration, and non-idealities of the real prototype.

For both, the corrected simulation reproduces the measured momentum dependence, and the residual difference is absorbed by a single, momentum-independent constant term. This shows that the framework reproduces the essential energy and time resolution behaviour of a real W-GAGG module across two decades in momentum. The modelling of the physical prototype is therefore adequate for the detector design studies for which the framework is intended.

\section{Summary and discussion}\label{sec:discussion}

The main outcome of the work presented here is the hybrid method itself. The optical transport, which dominates the CPU cost of a detailed simulation of an optical calorimeter, is factorised out of the event loop and replaced by a summary Monte Carlo step, driven by probability density functions that are extracted once, by full ray-tracing, from the module under study. The method has been developed and tested within the framework described in Section~\ref{sec:framework}, but it is not tied to it: it rests only on the assumption that the detector response can be split into an energy-deposition stage and an optical-transport stage.

The validation of the method was carried out entirely Monte Carlo to Monte Carlo, on one representative module per technology, and at three levels of increasing inclusiveness, namely the comparison of the photon time of arrival, of the photon yield, and of the reconstructed energy and time resolutions. At the most microscopic level, the arrival times of the individual optical photons at the photo-detector faces were compared pulse by pulse. The Kolmogorov-Smirnov test returns $p>0.05$ for more than $93\%$ of the pulses, in both technologies and in every readout channel and time window (Table~\ref{table:ks_summary}). %

At the level of the photon yield, the ratio between the number of photons extracted by HybridMC and by the full ray-tracing reference is centred close to unity in both technologies, with a slight positive bias of a few percent. The W-GAGG distributions are sharply peaked, while the Shashlik ones are visibly broader. This difference is most likely related to the optical calibration of the Shashlik technology. As discussed in Section~\ref{sec:opticalcalibration}, the Shashlik calibration must be performed independently for every scintillating tile, because the tiles are not equivalent to one another with respect to the WLS-fibre lattice and the module boundaries. Its grid can therefore not be made as dense as the per-fibre SpaCal one without inflating the calibration table beyond practical limits, and this coarser spatial sampling is the most plausible cause of the larger event-by-event variance of the parametrised photon yield. The effect is a broadening of the yield distribution rather than a shift of its centre, which is consistent with a sampling granularity that is too coarse rather than with a bias in the underlying parametrisation.

The third and most inclusive level is that of the reconstructed figures of merit, which are what the detector design actually depends on. For the W-GAGG module the two simulation approaches are equivalent within the fitted uncertainties. The sampling and constant terms of the energy resolution agree, and the time-resolution curves track each other over the whole $1$--$100$~GeV/c range, down to the $10$~ps level at the highest momenta. At this level the residual imperfections seen at the photon level do not propagate into any measurable difference.

For the Shashlik module the picture is more nuanced.
The energy resolution obtained with HybridMC is slightly but systematically worse than the full ray-tracing one at every momentum tested. This could be explained by the additional event-by-event spread of the number of extracted photons shown in Figure~\ref{fig:photon_ratio_shashlik}. The effect is nevertheless small, and the fitted parameters of the two curves, $a = (8.01 \pm 0.11)\%$ and $b = (0.73 \pm 0.04)\%$ for the full ray-tracing reference against $a = (8.47 \pm 0.12)\%$ and $b = (0.77 \pm 0.04)\%$ for HybridMC, remain reasonably compatible with each other within their statistical uncertainties. The energy resolution of the Shashlik module is therefore reproduced, with a systematic tendency of the parametrised transport to be marginally pessimistic.
The time resolution shows the opposite behaviour. HybridMC is systematically better than the full ray-tracing reference at every momentum point, and the difference grows with the primary-particle momentum, from about one percent at $1$~GeV/c to close to thirty percent at $100$~GeV/c. This could be correlated to the positive bias of a few percent in the number of extracted photons, since time resolution is driven by the photon statistics on the leading edge of the pulses. This effect is amplified at high momentum, where the resolution approaches the floor set by the scintillation rise time, by the single-photon time resolution of the photo-detector, and by the electronics noise.
For the Shashlik technology, therefore, the hybrid approach predicts at present a better time resolution than the full ray-tracing reference. This should be kept in mind whenever the framework is used to set timing specifications for Shashlik modules.

Both effects have the same probable root, namely the coarser spatial sampling of the Shashlik optical calibration discussed above. The natural way to remove them is to refine that sampling without inflating the calibration table beyond practical limits. A promising direction, left to future development, is a multi-step calibration of the Shashlik modules, in which the optical transport is factorised into successive stages, from the scintillating tile to the WLS fibre and from the fibre to the readout, so that each stage can be calibrated on its own grid and sampled at the granularity it requires. This would decouple the density of the spatial sampling from the overall size of the calibration table, which is what currently limits the Shashlik calibration.

The gain in CPU time is of more than two orders of magnitude in every configuration tested, ranging from $157\pm13$ for the W-GAGG SpaCal module to $481\pm80$ for the Pb-Polystyrene one (Table~\ref{table:speedup}). The spread between technologies is large and is worth examining.
The CPU cost of the parametrised step scales essentially with the number of optical photons that have to be generated and sampled, each of which costs a small and roughly fixed number of table look-ups. The cost of full ray-tracing, in contrast, scales with the number of photons multiplied by the mean number of optical interactions that each of them undergoes before being absorbed or extracted. To first order, therefore, the speed-up is governed by how long a photon survives inside the module before its fate is decided, and the quantities that set the absolute number of photons, namely the sampling fraction and the light yield of the scintillator, largely cancel in the ratio.

This picture is consistent with the measured numbers. The W-GAGG module has by far the most expensive parametrised step, $7.30$~s/GeV against $3$ to $4$~s/GeV for the other configurations, as expected from the high light yield of GAGG, which produces the largest number of photons to be sampled. Its full ray-tracing cost, however, is not correspondingly the highest, most likely because the slightly unpolished surfaces of the W-GAGG crystals scatter photons diffusely and terminate their histories comparatively early, making each individual photon cheap to track. The combination of an expensive parametrised step and a comparatively cheap ray-traced one yields the smallest speed-up of the four. At the other extreme, the cladded polystyrene fibres of the Pb-Polystyrene module trap photons in long sequences of total internal reflection before they are extracted or absorbed, which makes them the most expensive to ray-trace, at $2113$~s/GeV, and yields the largest speed-up. The interplay of light yield, sampling fraction, surface state and scintillator length is therefore the most plausible explanation of the observed ordering. Disentangling the individual contributions would require a dedicated study in which each is varied in turn.

The practical consequence of these speed-up factors is best appreciated at the scale of the full detector. Simulating the complete upgraded ECAL response to the minimum-bias particle flux expected at the HL-LHC Run~5 luminosity costs, with HybridMC, a mean of $(4.4\pm2.3)$~hours of CPU per event (Figure~\ref{fig:flux_cpu_time}). The same simulation performed with full ray-tracing of the optical photons would require tens of days per event, which is unfeasible in practice and would be extremely challenging even on a large distributed computing cluster. The parametrisation is thus what makes the detailed optical simulation of full-flux LHCb bunch crossings possible at all, rather than merely faster.

These results delineate the niche occupied by HybridMC within the landscape of accelerated calorimeter simulation. Full Geant4 ray-tracing remains the unavoidable reference, but it is impractical for any large-scale study of optical calorimeters at the LHCb Upgrade~II scale, for the reasons just given. At the opposite end of the spectrum, ultra-fast flash simulators based on generative deep-learning models, such as Lamarr~\cite{LHCb_Lamarr_2024}, achieve speed-ups of five orders of magnitude by parametrising the macroscopic detector response. This approach is ideal for high-statistics physics analyses, but by construction it cannot expose the internal optical response of the detector and is therefore unsuitable for the design-iteration phase, where modifications of the surface state, of the absorber geometry, or of the readout coupling must be propagated to the photon yield and the timing distribution. Hardware-accelerated ray-tracing toolkits such as Opticks~\cite{Opticks} or Mitsuba3~\cite{davis2023mitsuba3} retain microscopic fidelity but require GPU resources. HybridMC fills the gap between these two regimes: it preserves the full microscopic optical physics of a Geant4 simulation by encoding it once into the calibration probability density functions, and then re-uses this information at CPU cost compatible with large-scale R\&D campaigns on existing computing infrastructures.

The comparison to test-beam data reported in Section~\ref{sec:comparison}
establishes whether the description of a real module implemented in the framework, that is its geometry, its materials, its surface states and its optical properties, is faithful enough to reproduce the behaviour of the physical object. For the W-GAGG prototype this is largely the case: the framework reproduces the momentum dependence of both the measured energy resolution and the measured time resolution over the two decades covered by the DESY~II and SPS campaigns, up to a small, momentum-independent residual that a fitted constant term absorbs. %
Residual discrepancies at this level, in this or in any other module, would point to an inaccuracy in the description of that specific module or of the experimental setup, and not to a shortcoming of the hybrid method, which is independent of the module configuration.

The scope of the parametrisation is bounded by its underlying assumptions. The calibration probability density functions are valid only for the specific module configuration on which they have been produced: any change in the absorber geometry, fibre lattice, surface roughness, or in the optical properties of the reflectors and the cladding materials requires a new calibration to be generated. The same applies to radiation-damage studies, in which the wavelength-dependent absorption length of the active medium is modified to emulate the accumulated dose~\cite{collaboration2021framework}.

The framework has been exercised on every SpaCal and Shashlik configuration relevant to the LHCb ECAL Upgrade~II programme, has produced the full Run~4 and Run~5 ECAL Monte Carlo samples~\cite{LHCb_TDR_023,LHCb_TDR_024} and supported dedicated studies of the calorimeter performance and readout~\cite{Roux2025,Govorkova_2026}, and is currently the reference fast-detailed simulation tool for the optical optimisation of the upgraded calorimeter. The same architecture is in principle applicable to any optical calorimeter or scintillator-based detector whose response can be factorised into an energy-deposition stage and an optical-transport stage, and adaptations to detectors outside the LHCb context are foreseen. The calorimeters proposed for the Future Circular Collider (FCC) and, more generally, the optical-calorimeter technologies developed within the DRDCalo collaboration~\cite{DRD6_proposal} are natural candidates. The dual-readout calorimeters foreseen for FCC-ee collect the scintillation and the Cherenkov light of a shower separately, in order to reconstruct its electromagnetic fraction event by event. They are pursued both in a fibre-sampling geometry, with scintillating and clear fibres embedded in a dense absorber, which shares the fibre-in-absorber layout of the SpaCal modules described here, and in a homogeneous-crystal geometry~\cite{DRD6_dro_crystals}. Radiation-hard sampling designs such as RADiCAL, built from alternating dense-absorber and scintillating-crystal plates read out by wavelength-shifting fibres~\cite{DRD6_radical}, share instead the longitudinally segmented layout of the Shashlik modules. Finer-sampling concepts such as the GRAiNITA grain calorimeter, in which millimetric high-density scintillator grains are immersed in a transparent heavy liquid and read out by wavelength-shifting fibres~\cite{DRD6_grainita}, push the optical complexity further still. In all of these the detector response is governed by the transport of a large number of optical photons, the same combination of microscopic optical detail and prohibitive photon multiplicity that motivated the development of HybridMC.

\section{Conclusions}\label{sec:conclusions}

The HybridMC framework provides a fast detailed Monte Carlo simulation of optical calorimeters by decoupling the high-energy shower development, handled by Geant4, from the optical photon transport, reproduced via a parametrised summary Monte Carlo step driven by probability density functions extracted from a one-time full ray-tracing optical calibration. A Monte Carlo-to-Monte Carlo validation has shown statistical equivalence of the output with full ray-tracing, with the Kolmogorov-Smirnov test confirming good agreement for more than $93\%$ of the simulated photo-detector pulses in the validated configurations. The framework yields acceleration factors ranging from $\sim$$150$ for W-GAGG SpaCal modules to $\sim$$480$ for Pb-Polystyrene SpaCal modules, making the detailed simulation of full-flux LHCb bunch crossings computationally feasible. The energy and time resolutions measured in test beams at DESY~II and at the CERN SPS on a W-GAGG SpaCal prototype are reproduced by the simulation up to a small residual constant term, demonstrating that the description of a real module implemented in the framework is adequate. HybridMC is currently the reference fast detailed simulation tool for the optical optimisation of the LHCb ECAL Upgrade~II, and its modular architecture makes it directly applicable to any optical calorimeter or scintillator-based detector whose response can be factorised into an energy-deposition stage and an optical-transport stage.

\section*{Software availability}\label{sec:availability}
The HybridMC source code is maintained within the LHCb collaboration infrastructure and is currently available to members of the LHCb collaboration. Access for external users can be granted on request. %
The framework is distributed together with the configuration files reproducing a few relevant module geometries. %

\section*{Acknowledgements}\label{sec:ack}
The authors thank Andreas Schopper and Yuri Guz for fruitful discussions. We are grateful to our colleagues from the LHCb Simulation Project~\cite{CERN-LHCC-2018-007} for enabling the production of the simulated minimum-bias samples used in Section~\ref{sec:gain}. This work made use of computing resources provided by CERN Lxplus.

\section*{Declaration of generative AI and AI-assisted technologies in the manuscript preparation process}

During the preparation of this work, the author(s) used ClaudeAI for language refinement and formatting assistance. The author(s) reviewed and edited the output as needed and take full responsibility for the content of the published article.

\bibliographystyle{elsarticle-num}         %

\bibliography{cas-refs}

\begin{thebibliography}{10}
\expandafter\ifx\csname url\endcsname\relax
  \def\url#1{\texttt{#1}}\fi
\expandafter\ifx\csname urlprefix\endcsname\relax\def\urlprefix{URL }\fi
\expandafter\ifx\csname href\endcsname\relax
  \def\href#1#2{#2} \def\path#1{#1}\fi

\bibitem{The_LHCb_Collaboration_2008}
A.~A. Alves, et~al., \href{https://doi.org/10.1088/1748-0221/3/08/S08005}{The
  lhcb detector at the lhc}, Journal of Instrumentation 3~(08) (2008) S08005.
\newblock \href {https://doi.org/10.1088/1748-0221/3/08/S08005}
  {\path{doi:10.1088/1748-0221/3/08/S08005}}.
\newline\urlprefix\url{https://doi.org/10.1088/1748-0221/3/08/S08005}

\bibitem{calo1_2000}
S.~Amato, et~al., \href{https://cds.cern.ch/record/494264}{Lhcb calorimeters:
  Technical design report}, Tech. Rep. CERN-LHCC-2000-036, CERN (2000).
\newline\urlprefix\url{https://cds.cern.ch/record/494264}

\bibitem{calo2_2020}
C.~A. Beteta, et~al., \href{https://cds.cern.ch/record/2729028}{Calibration and
  performance of the lhcb calorimeters in run 1 and 2 at the lhc}, Tech. Rep.
  LHCb-DP-2020-001, CERN (2020).
\newline\urlprefix\url{https://cds.cern.ch/record/2729028}

\bibitem{SpaCal_proposal_1987}
P.~Jenni, P.~Sonderegger, H.~P. Paar, R.~Wigmans, {The High Resolution
  Spaghetti Hadron Calorimeter: Proposal}, Tech. Rep. NIKHEF-H/87-7, NIKHEF,
  Amsterdam (1987).

\bibitem{collaboration2021framework}
{LHCb Collaboration}, Framework tdr for the lhcb upgrade ii: Opportunities in
  flavour physics, and beyond, in the hl-lhc era cern-lhcc-2021-012, Tech.
  rep., LHCB-TDR (2021).

\bibitem{agostinelli2003geant4}
S.~Agostinelli, et~al., Geant4 -- a simulation toolkit, Nuclear Instruments and
  Methods in Physics Research Section A 506 (2003) 250--303.
\newblock \href {https://doi.org/10.1016/S0168-9002(03)01368-8}
  {\path{doi:10.1016/S0168-9002(03)01368-8}}.

\bibitem{LHCb_Lamarr_2024}
L.~Anderlini, et~al., \href{https://doi.org/10.1051/epjconf/202429503040}{The
  lhcb ultra-fast simulation option, lamarr design and validation}, EPJ Web of
  Conferences 295~(03040) (may 2024).
\newblock \href {https://doi.org/10.1051/epjconf/202429503040}
  {\path{doi:10.1051/epjconf/202429503040}}.
\newline\urlprefix\url{https://doi.org/10.1051/epjconf/202429503040}

\bibitem{cds_fast_simulation_2021}
{LHCb Collaboration}, \href{https://cds.cern.ch/record/2781378}{Performance of
  the fast simulation interface in gauss-on-gaussino}, Tech. Rep.
  LHCB-FIGURE-2021-004, CERN (2021).
\newline\urlprefix\url{https://cds.cern.ch/record/2781378}

\bibitem{cds_michal_2021}
M.~Mazurek, G.~Corti, D.~Müller, \href{https://cds.cern.ch/record/2790591}{New
  simulation software technologies at the lhcb experiment at cern}, Tech. Rep.
  LHCb-PROC-2021-011, CERN (2021).
\newline\urlprefix\url{https://cds.cern.ch/record/2790591}

\bibitem{Levin1996Unified}
A.~Levin, C.~Moisan, A more physical approach to model the surface treatment of
  scintillation counters and its implementation into {DETECT}, in: IEEE Nuclear
  Science Symposium Conference Record, Vol.~2, 1996, pp. 702--706.
\newblock \href {https://doi.org/10.1109/NSSMIC.1996.591410}
  {\path{doi:10.1109/NSSMIC.1996.591410}}.

\bibitem{cds_michal_2025}
M.~Mazurek, \href{https://cds.cern.ch/record/2945115}{Machine learning in lhcb
  simulation: From fast to flash}, Tech. Rep. LHCb-PROC-2025-004, CERN (2025).
\newline\urlprefix\url{https://cds.cern.ch/record/2945115}

\bibitem{chekalina2019generative}
V.~Chekalina, E.~Orlova, F.~Ratnikov, D.~Ulyanov, A.~Ustyuzhanin, E.~Zakharov,
  Generative models for fast calorimeter simulation: the lhcb case, in: EPJ Web
  of Conferences, Vol. 214, EDP Sciences, 2019, p. 02034.

\bibitem{Opticks}
{Blyth, Simon}, \href{https://doi.org/10.1051/epjconf/202024511003}{Meeting the
  challenge of juno simulation with opticks: Gpu optical photon acceleration
  via nvidia optixtm}, EPJ Web Conf. 245 (2020) 11003.
\newblock \href {https://doi.org/10.1051/epjconf/202024511003}
  {\path{doi:10.1051/epjconf/202024511003}}.
\newline\urlprefix\url{https://doi.org/10.1051/epjconf/202024511003}

\bibitem{davis2023mitsuba3}
A.~C.~S. Davis, S.~Barré, Y.~Cui, K.~L. Evans, M.~Gersabeck, A.~Rat,
  Z.~Montazeri, \href{https://arxiv.org/abs/2309.12496}{Optical photon
  simulation with mitsuba3} (2023).
\newblock \href {http://arxiv.org/abs/2309.12496} {\path{arXiv:2309.12496}}.
\newline\urlprefix\url{https://arxiv.org/abs/2309.12496}

\bibitem{An_2022_TB}
L.~An, et~al., \href{https://doi.org/10.1016/j.nima.2022.167629}{Performance of
  a spaghetti calorimeter prototype with tungsten absorber and garnet crystal
  fibres}, Nuclear Instruments and Methods in Physics Research Section A 1045
  (2023) 167629.
\newblock \href {https://doi.org/10.1016/j.nima.2022.167629}
  {\path{doi:10.1016/j.nima.2022.167629}}.
\newline\urlprefix\url{https://doi.org/10.1016/j.nima.2022.167629}

\bibitem{CERN-LHCC-2000-036}
{LHCb Collaboration}, \href{https://cds.cern.ch/record/494264}{{LHCb
  calorimeters : Technical Design Report}}, Tech. rep., CERN, Geneva (2000).
\newline\urlprefix\url{https://cds.cern.ch/record/494264}

\bibitem{Birks1951}
J.~B. Birks, Scintillations from organic crystals: specific fluorescence and
  relative response to different radiations, Proceedings of the Physical
  Society. Section A 64~(10) (1951) 874--877.
\newblock \href {https://doi.org/10.1088/0370-1298/64/10/303}
  {\path{doi:10.1088/0370-1298/64/10/303}}.

\bibitem{Badhwar1967}
G.~D. Badhwar, C.~L. Deney, B.~R. Dennis, M.~F. Kaplon, The non-linear response
  of the plastic scintillator {NE102}, Nuclear Instruments and Methods 57
  (1967) 116--120.
\newblock \href {https://doi.org/10.1016/0029-554X(67)90664-2}
  {\path{doi:10.1016/0029-554X(67)90664-2}}.

\bibitem{GAGG_Birks}
M.~Kole, N.~De~Angelis, {Measurements of the Birks' coefficient of GAGG:Ce
  using hard X-rays}, Nuclear Instruments and Methods in Physics Research
  Section A 1084 (2026) 171206.
\newblock \href {https://doi.org/10.1016/j.nima.2025.171206}
  {\path{doi:10.1016/j.nima.2025.171206}}.

\bibitem{Ritt2008DRS4}
S.~Ritt, R.~Dinapoli, U.~Hartmann, Application of the {DRS} chip for fast
  waveform digitizing, Nuclear Instruments and Methods in Physics Research
  Section A 623~(1) (2010) 486--488.
\newblock \href {https://doi.org/10.1016/j.nima.2010.03.045}
  {\path{doi:10.1016/j.nima.2010.03.045}}.

\bibitem{PDG2024}
S.~Navas, et~al., {Review of Particle Physics}, Phys. Rev. D 110~(3) (2024)
  030001.
\newblock \href {https://doi.org/10.1103/PhysRevD.110.030001}
  {\path{doi:10.1103/PhysRevD.110.030001}}.

\bibitem{CERN-LHCC-2018-007}
{LHCb Collaboration}, \href{https://cds.cern.ch/record/2310827}{{Upgrade
  Software and Computing}}, Tech. rep., CERN, Geneva (2018).
\newblock \href {https://doi.org/10.17181/CERN.LELX.5VJY}
  {\path{doi:10.17181/CERN.LELX.5VJY}}.
\newline\urlprefix\url{https://cds.cern.ch/record/2310827}

\bibitem{Roux_GAGG_PicoCal}
L.~Roux, et~al., {Development of Ultrafast, Radiation-Hard GAGG Scintillators
  for the Upgrade II Electromagnetic Calorimeter of LHCb: PicoCal}, IEEE
  Transactions on Nuclear Science (submitted) (2025).

\bibitem{LHCb_TDR_023}
{LHCb Collaboration}, \href{https://cds.cern.ch/record/2903094}{Lhcb upgrade ii
  scoping document, cern-lhcc-2024-008, lhcb-tdr-023}, Tech. rep., CERN (2024).
\newline\urlprefix\url{https://cds.cern.ch/record/2903094}

\bibitem{LHCb_TDR_024}
{LHCb Collaboration}, \href{https://cds.cern.ch/record/2903095}{Lhcb ecal
  enhancement technical design report, cern-lhcc-2024-009, lhcb-tdr-024}, Tech.
  rep., CERN (2024).
\newline\urlprefix\url{https://cds.cern.ch/record/2903095}

\bibitem{Roux2025}
L.~Roux, L.~Martinazzoli, M.~Pizzichemi, C.~Dujardin, E.~Auffray, Influence of
  crystal fiber inhomogeneity on the energy resolution of a sampling
  electromagnetic calorimeter, IEEE Transactions on Nuclear Science 72~(7)
  (2025) 2076--2081.
\newblock \href {https://doi.org/10.1109/TNS.2025.3559721}
  {\path{doi:10.1109/TNS.2025.3559721}}.

\bibitem{Govorkova_2026}
K.~Govorkova, J.~García~Pardiñas, V.~Lončar, S.~Schmitt, V.~Nguyen,
  M.~Pizzichemi, L.~Martinazzoli, E.~Anne~Smith,
  \href{https://doi.org/10.1088/2632-2153/ae8218}{Enabling low-latency machine
  learning on radiation-hard fpgas with hls4ml}, Machine Learning: Science and
  Technology 7~(4) (2026) 045024.
\newblock \href {https://doi.org/10.1088/2632-2153/ae8218}
  {\path{doi:10.1088/2632-2153/ae8218}}.
\newline\urlprefix\url{https://doi.org/10.1088/2632-2153/ae8218}

\bibitem{DRD6_proposal}
{DRD6 Collaboration}, \href{https://cds.cern.ch/record/2886494}{{DRD6:
  Calorimetry}}, Tech. Rep. CERN-DRDC-2024-004, DRDC-P-DRD6, CERN, Geneva
  (2024).
\newline\urlprefix\url{https://cds.cern.ch/record/2886494}

\bibitem{DRD6_dro_crystals}
R.~Hirosky, et~al., \href{https://arxiv.org/abs/2408.11973}{{Dual-readout
  calorimetry with homogeneous crystals}}, EPJ Web Conf. (CALOR2024
  proceedings) (2024).
\newblock \href {http://arxiv.org/abs/2408.11973} {\path{arXiv:2408.11973}}.
\newline\urlprefix\url{https://arxiv.org/abs/2408.11973}

\bibitem{DRD6_radical}
{RADiCAL Collaboration}, \href{https://arxiv.org/abs/2401.01747}{{Study of time
  and energy resolution of an ultra-compact sampling calorimeter (RADiCAL)
  module at EM shower maximum over the energy range 25~GeV $\le$ E $\le$
  150~GeV}} (2024).
\newblock \href {http://arxiv.org/abs/2401.01747} {\path{arXiv:2401.01747}}.
\newline\urlprefix\url{https://arxiv.org/abs/2401.01747}

\bibitem{DRD6_grainita}
S.~Barsuk, et~al., {First characterization of a novel grain calorimeter: the
  GRAiNITA prototype}, JINST 19 (2024) P04008.
\newblock \href {http://arxiv.org/abs/2312.07365} {\path{arXiv:2312.07365}},
  \href {https://doi.org/10.1088/1748-0221/19/04/P04008}
  {\path{doi:10.1088/1748-0221/19/04/P04008}}.

\end{thebibliography}

\end{document}